\documentclass[useAMS,usenatbib]{mn2e}
\usepackage[totalwidth=169mm, totalheight=243mm]{geometry}
\usepackage[dvips]{graphicx}
\usepackage{pslatex}
\usepackage{amssymb,amsmath}
\usepackage{color,psfrag}
\usepackage{verbatim,xspace}
\usepackage{import,afterpage}
\usepackage{supertabular}
\usepackage[varg]{txfonts}

\newcommand{\ro}[1]{\ensuremath{\textrm{#1}}}
\newcommand{\kmsec}{\ensuremath{~\ro{km}~ \ro{s}^{-1}}\xspace}
\newcommand{\ergs}{\ensuremath{~\ro{erg s}^{-1}}}
\newcommand{\cm}{\ensuremath{~\ro{cm}^{-2}}}
\newcommand{\Msol}{\ensuremath{M_{\odot}}\xspace}
\newcommand{\lya}{\ensuremath{\ro{Ly}\alpha}\xspace}

\newcommand{\hi}{\ensuremath{\ro{H\textsc{i}}}\xspace}

\newcommand{\df}{\ensuremath{~ \ro{d} }}
\newcommand{\dd}{\ensuremath{\ro{d} }}
\newcommand{\rv}{\ensuremath{\boldsymbol{r}}\xspace}
\newcommand{\vv}{\ensuremath{\boldsymbol{v}}\xspace}

\newcommand{\gad}{\textsc{gadget}-2\xspace}
\newcommand{\cN}{\ensuremath{\mathcal{N}}\xspace}

\defcitealias{2008ApJ...681..856R}{R08}
\defcitealias{2009MNRAS.397..511B}{BH09}
\defcitealias{2010MNRAS.403..870B}{BH10}

\newcommand{\imgdir}{.}
\newcommand{\mnpg}[1]{\begin{minipage}[c]{0.41\textwidth} \centering
	#1 \end{minipage} }
\newcommand{\nudge}[1]{\begin{minipage}[c]{#1\textwidth} \centering \end{minipage} }	
\newcommand{\vsp}{\vspace{-0.4cm}}
\newcommand{\hsp}{\hspace{1.0cm}}

\newcommand{\impth}[2]{\imgdir/#1_grid#2_} 

\newcommand{\pmcomp}[5]{ 
	\begin{figure*} \centering
		\vspace{-0.3cm} 
		\nudge{0.01}\hspace{-0.21cm}\mnpg{\import{\impth{#1}{#2}}{Nim#2#31.eps}} \hsp \vspace{-0.92cm}
		\mnpg{\import{\impth{#1}{#2}}{Nim#2#32.eps}} 
		\nudge{0.01}\hspace{0.5cm}\mnpg{\import{\impth{#1}{#2}}{Sim#2#31.eps}} \hsp 
		\mnpg{\import{\impth{#1}{#2}}{Sim#2#32.eps}} 
		\nudge{1}
		\nudge{0.01}\mnpg{\import{\impth{#1}{#2}}{Ssp#2#311v.eps}} \hsp \vspace{-0.97cm}
		\mnpg{\import{\impth{#1}{#2}}{Ssp#2#321v.eps}} 
		\nudge{0.005}\mnpg{\import{\impth{#1}{#2}}{Ssp#2#312v.eps}} \hsp 
		\mnpg{\import{\impth{#1}{#2}}{Ssp#2#322v.eps}} 
	
		\caption{#4} \label{#5}
	\end{figure*}
	}

\newcommand{\windcomp}[7]{ 
	\begin{figure*} \centering
		\vspace{-0.3cm} 
		\nudge{0.01}\hspace{-0.21cm}\mnpg{\import{\impth{#1}{#3}}{Nim#3#4#5.eps}} \hsp \vspace{-0.92cm}
		\mnpg{\import{\impth{#2}{#3}}{Nim#3#4#5.eps}} 
		\nudge{0.01}\hspace{0.5cm}\mnpg{\import{\impth{#1}{#3}}{Sim#3#4#5.eps}} \hsp 
		\mnpg{\import{\impth{#2}{#3}}{Sim#3#4#5.eps}} 
		\nudge{1}
		\nudge{0.01}\mnpg{\import{\impth{#1}{#3}}{Ssp#3#4#51v.eps}} \hsp \vspace{-0.97cm}
		\mnpg{\import{\impth{#2}{#3}}{Ssp#3#4#51v.eps}} 
		\nudge{0.005}\mnpg{\import{\impth{#1}{#3}}{Ssp#3#4#52v.eps}} \hsp 
		\mnpg{\import{\impth{#2}{#3}}{Ssp#3#4#52v.eps}} 
	
		\caption{#6} \label{#7}
	\end{figure*}
	}

\title[Galactic Winds and Extended \lya Emission]{Galactic Winds and Extended \lya 
Emission from the Host Galaxies of High Column Density QSO Absorption Systems}

\author[Barnes, Haehnelt, Tescari and Viel]{Luke A. Barnes$^{1}$
\thanks{E-Mail: luke.barnes@phys.ethz.ch (LAB)}, Martin G. Haehnelt$^{2}$\footnotemark[1],
Edoardo Tescari$^{3,4,5}$ and Matteo Viel$^{3,4}$ \\
$^1$Institute for Astronomy, ETH Zurich, Wolfgan-Pauli-Strasse 27, CH-8093, Zurich\\
$^2$Institute of Astronomy and Kavli Institute for Cosmology, Madingley Road, Cambridge, CB3
0HA \\
$^3$INAF - Osservatorio Astronomico di Trieste, Via G.B. Tiepolo 11, I-34131 Trieste, Italy\\
$^4$INFN/National Institute for Nuclear Physics, Via Valerio 2, I-34127 Trieste, Italy\\
$^5$Dipartimento di Fisica - Sezione di Astronomia, Universit\'a di Trieste, 
Via G.B. Tiepolo 11, I-34131 Trieste, Italy\\
}

\begin{document}

\date{not yet submitted}

\pagerange{\pageref{firstpage}--\pageref{lastpage}} \pubyear{2010}

\maketitle 

\label{firstpage}

\begin{abstract} 
We present 3D resonant radiative transfer simulations 
of the spatial and spectral diffusion of the \lya radiation 
from a central source in the host galaxies of high column density
absorption systems at $z\sim 3$. The radiative transfer simulations
are based on a suite of 
cosmological galaxy formation simulations which 
reproduce a wide range of observed properties of 
damped \lya absorption systems. The \lya emission is predicted to be 
spatially extended up to several arcsec, and the spectral width of
the \lya emission is broadened to several hundred (in some case
more than  thousand) \kmsec. The distribution and the dynamical state 
of the gas in the simulated galaxies is complex, the latter 
with significant contributions from rotation and both in- and
out-flows. The emerging \lya radiation extends to gas with column
densities of $N_\hi \sim 10^{18} \cm$ and its spectral shape varies 
strongly with viewing angle. 
The strong dependence on the central \hi column density and the \hi 
velocity field suggests that the \lya emission will also vary strongly with time
on timescales of a few dynamical times of the central
region. Such variations with time should be  especially pronounced 
at times where the host galaxy  undergoes a  major merger and/or starburst. 
Depending on the pre-dominance of
in- or out-flow along a given sightline and the
central column density, the spectra show prominent blue peaks, red
peaks or double-peaked profiles. Both spatial distribution and 
spectral shape are very sensitive to details of the galactic wind 
implementation. Stronger galactic winds result in more spatially
extended \lya emission and -- somewhat counterintuitively -- 
a narrower spectral distribution. 
\end{abstract}

\begin{keywords} quasars: absorption lines --- galaxies: formation
\end{keywords}


\section{Introduction} 

Searches for \lya emission that are performed 
down to very low flux levels can detect very faint objects 
where the \lya emission is little affected by dust.
\lya emission has therefore developed into an important tool 
to study high-redshift galaxies \citep{1998AJ....115.1319C,2000ApJ...532..170S,2002ApJ...568L..75H,
2003PASJ...55L..17K,2004ApJ...611...59R,2005PASJ...57..165T,2006ApJ...648....7K,
2006Natur.443..186I,2007MNRAS.376..727S,2008ApJ...677...12O,2010arXiv1007.2961O}.
At high redshift, or for 
intrinsically faint objects, it is also often the only means of establishing 
a spectroscopic redshift. At low flux levels, \lya emission due to 
star formation in galaxies can be spatially
very extended due to the resonant scattering of \lya 
photons --- even if \lya cooling is not important and 
the corresponding continuum emission is very compact 
\citep{2004AJ....128.2073H,2006ApJ...649...14D,2008ApJ...681..856R,2010arXiv1008.0634F}. 

Deep \lya observations of the high-redshift Universe thus 
have tremendous potential to probe the surrounding neutral gas in
protogalaxies, and thereby shed light on galaxy 
formation. The spectroscopic survey of \citet[][hereafter
\citetalias{2008ApJ...681..856R}]{2008ApJ...681..856R} for low surface
brightness \lya emitters based on a 92 hour long exposure with the ESO
VLT FORS2 instrument has pushed \lya surveys to new sensitivity limits
and yielded a sample of 27 spatially extended (several arcsec) low-surface
\lya emitters with fluxes of $3\times 10^{-18} {\rm erg} {\rm s}^{-1}
{\rm cm}^{-1}$ and a space density of $\sim 3\times 10^{-2}h_{70}^{3}{\rm
 Mpc}^{-3}$, similar to that of the faintest dwarf galaxies in the
local Universe. Based on the inferred incidence rate for absorption, 
\citetalias{2008ApJ...681..856R}
argued that their sample should overlap significantly
with the elusive host population of Damped \lya
Absorption Systems (DLAs). 

\citet[][hereafter \citetalias{2009MNRAS.397..511B}, \citetalias{2010MNRAS.403..870B}]
{2009MNRAS.397..511B,2010MNRAS.403..870B}, building on the successful model for DLAs of
\citet{1998ApJ...495..647H,2000ApJ...534..594H}, presented a simple
model that simultaneously accounts for the kinematic properties,
column density distribution and incidence rate of DLAs \emph{and} the
luminosity function and the size distribution of the
\citetalias{2008ApJ...681..856R} emitters in the context of the
$\Lambda$CDM model for structure
formation. The modelling of \citetalias{2010MNRAS.403..870B} assumed spherical
symmetry and simple power law
profiles for gas density and peculiar velocity. 
\lya radiative transfer is, however, sensitive to the 
details of the spatial distribution and velocity of neutral hydrogen ---
inhomogeneity can provide channels for rapid spatial diffusion, while
bulk velocities have a similar effect in frequency space.

The complicated physics of galaxy formation --- gravity,
hydrodynamics, radiative cooling, star formation, supernovae, galactic
winds, ionising radiation, metal and dust production --- has led to
numerical simulations being the method of choice for investigating 
the physical properties of the gas from which galaxies form. These simulations provide us
with realistic 3D density and velocity fields that take into account
most of the relevant physics.

Here we will apply the \lya radiative transfer code developed in
\citetalias{2010MNRAS.403..870B}, adapted for the post-processing of 
3D gas distributions, to the results of simulation of DLA host
galaxies in a cosmological context by \citet{2009MNRAS.397..411T}.
We study the effect of the spatial distribution and the kinematic 
state of the neutral hydrogen on the low surface brightness 
spatially extended \lya emission, which should become 
an important probe of kinematics of the gas in DLA host
galaxies.

In this first paper we will focus on a detailed investigation 
of a small number of such simulated systems. 
  In Section \ref{S:lya3D}, we describe the (minor) changes to our \lya radiative
transfer code needed to follow photons through a density
distribution on a 3D grid. 
  In Section \ref{S:tesc}, we describe the set of simulations of
\citet{2009MNRAS.397..411T} that we have used, and in 
Section \ref{S:3haloes} we describe the three haloes that 
we have extracted from the simulations for further study.
  Section \ref{S:galwind} discusses the haloes in detail,
looking particularly at the effect of galactic  winds on the 
gas properties and \lya emission.
  We discuss our results in Section \ref{S:discuss}, including 
their relevance to observed populations of high redshift \lya
emitters, including LBG's, before presenting our conclusions 
in Section \ref{S:conclusion}.


\section{The 3D \lya Radiative Transfer code} \label{S:lya3D} 

3D \lya cosmological radiative transfer of \lya radiation 
has been studied by a number of authors \citep{2005ApJ...628...61C,
2006ApJ...645..792T,2009ApJ...696..853L,2009arXiv0910.2712Z,
2010ApJ...708.1048K,2010ApJ...725..633F},
in the context of, for example, fluorescent emission from the IGM,
bright \lya emitters at $z \approx 8$, young Lyman break galaxies and
\lya emission linked to radiative cooling. 

We have previously described our implementation of a 1D \lya radiative
transfer algorithm in \citetalias{2010MNRAS.403..870B}. 
In short, \lya photons are created according to the emissivity profile, and 
then propagated through the gas. Each time the photon is scattered, a new frequency 
is calculated by assuming that the photon's frequency in the scattering atom's frame 
is unchanged. Thus, the code tracks the random walk in both real and 
frequency space before the photon escapes and is observed.
The new code used here follows photons through a regular 3D grid. 
The acceleration scheme, by which 
photons which have frequencies close to line centre are scattered into the wings, 
is adapted to the conditions in the current cell. We have 
tested and subsequently adopted a prescription similar to that described 
in \citet{2009ApJ...696..853L}. Further details on the code can be 
found in \citet{Barnesthesis}.

The emergent spectrum will depend on the angle from which the system
is viewed. Rather than ``waiting'' 
for enough photons to emerge in a given
direction, we implement the ``peeling-off'' algorithm, 
described in \citet{1984ApJ...278..186Y} and 
\citet{1999ApJ...525..799W}. At each scattering, the probability of
escape in the direction of the observer is calculated and a suitable
weight added to the corresponding 2D pixel. 
The photons that eventually escape the system are used to
calculate the angularly-averaged spectrum and surface brightness
profile. We have run standard test problems to test our 3D 
code and found that they are reproduced very well.

Our modelling nevertheless has a number of significant simplifications. 
We post-process the output of a cosmological simulations. The simulations 
therefore do not take into account the dynamical effect of radiation pressure. \lya 
radiation pressure has been proposed to provide sufficient energy to launch 
an outflowing supershell for luminous sources
\citep{2009MNRAS.396..377D}. However, the systems that we consider 
here do not have the necessary \lya luminosity for radiation pressure 
to be dynamically significant. Our simulations also ignore the evolution of the 
density and velocity field over the light travel time of the \lya photons.
We further only consider a very centrally concentrated 
emissivity. For simplicity we inject photons at the centre
of mass of the halo. The photons are created with line-centre
frequency in the fluid frame of the gas. We will discuss these assumptions 
further in Section \ref{S:limit}.


\section{\lya radiative transfer in DLA host galaxies } \label{S:RT}

\subsection{The Simulations of DLA host galaxies in a cosmological
context by Tescari et al.} \label{S:tesc}

We will make here use of the cosmological hydrodynamic simulations described in 
\citet{2009MNRAS.397..411T}. These simulations were aimed at
reproducing the physical properties of the host galaxies of DLAs at $z
\sim 3$. The numerical code is a modification of \gad, which is a parallel
Tree-PM SPH code \citep{2005MNRAS.364.1105S}. The code is extensively 
described in \citet{2007MNRAS.382.1050T}, where it was used to 
simulate clusters of galaxies; see also \citet{2010MNRAS.402.1911T} and 
\citet{2010arXiv1007.1628T}. In addition to gravity and hydrodynamics, the other 
physical processes that are modelled in the simulation are:

\begin{itemize}
\item Radiative cooling and heating, including a UV background produced by
quasars and galaxies.
\item Chemical evolution, using the model of \citet{2007MNRAS.382.1050T} 
which traces the following elements: H, He, C, O,
Mg, S, Si and Fe. The contribution of metals is included in the
cooling function. The release of metals from Type Ia and Type II
supernovae, as well as low- and intermediate-mass stars, is followed.
\item Star formation, using a multiphase criterion. 
\item An effective model for the ISM, whereby gas particles are treated 
as multiphase once they exceed a density threshold for the formation of 
cold clouds. These cold clouds are the precursors to stars. The neutral
hydrogen fraction depends on the UV background in the low density gas, and on 
the fraction $f_c$ of mass in cold clouds above the density threshold.
\item Galactic winds, using two different prescriptions. An
energy-driven wind model is implemented in the form of a fixed velocity kick
given to a chosen particle, with a mass-loss rate proportional to the
star formation rate. \citet{2009MNRAS.397..411T} also
investigated a
momentum-driven wind model that mimics a scenario in which the
radiation pressure of a starburst drives an outflow. In this model,
the velocity kick scales with the velocity dispersion of the
galaxy. These wind prescriptions are admittedly rather crude, and rely
on phenomenological parameters that are poorly constrained either by
observations or by more sophisticated modelling. This reflects the
significant uncertainty regarding the influence of winds on galaxy
formation.
\end{itemize}

\citet{2009MNRAS.397..411T} investigated simulations with varying box size and numerical
resolution for a range of different implementations of galactic
winds. They chose the ``SW'' (Strong Wind)
simulation with a box size of $10 h^{-1}$Mpc and $320^3$ particles, with a mass resolution
of $3.5 \times 10^5 h^{-1}\Msol$ as their fiducial simulation. This simulation employed a strong
(600 \kmsec), energy-driven wind and a Salpeter stellar IMF. Looking ahead to 
a favourable comparison with the \citetalias{2008ApJ...681..856R} 
emitters in a later section, we will 
instead focus more on the Momentum Driven Wind (MDW) simulation, 
which differs from the SW simulation only
in the wind implementation. We will compare the results of 
both the MDW and SW simulation with two further simulations: a weak (energy
driven, 100 \kmsec) wind (WW) and a simulation with no wind (NW) which did not 
appear in \citet{2009MNRAS.397..411T}. As we will see, the MDW and the SW simulations 
are somewhat similar, as are the WW and NW simulations.

The simulations of \citet{2009MNRAS.397..411T} reproduce most observed properties of DLAs rather 
well. The observed incidence rate of DLAs is matched by the
simulations, assuming that haloes below $10^9 h^{-1} \Msol$ do not
host DLAs. The observed column density distribution is also reproduced
successfully for the SW and MDW simulations, while the total neutral
gas mass in DLAs ($\Omega_\ro{DLA}$) is reproduced for MDW but underpredicted 
by a factor of about 1.7 at $z = 3$ in the SW simulation.
The most problematic observable is 
the velocity width distribution of low-ionization associated metal 
absorption. The high velocity tail of the distribution is
significantly underpredicted in all simulations. The simulations do
not produce enough absorption systems with velocity width greater than
100 \kmsec. This is a common problem for numerical simulations 
of DLA host galaxies \citep{2008MNRAS.390.1349P,2008ApJ...683..149R}, 
but see \citet{2010arXiv1008.4242H} and \citet{2010arXiv1010.5014C}
which appear to be somewhat more successful in this respect. We will
come back to this point later. 

\begin{table*}
\begin{tabular}{|c|c|c|c|c|c|c|c|} \hline Halo & Wind & Mass 	& Mass
\hi & Virial Velocity	& \multicolumn{2}{c|}{Virial Radius} & \lya
Luminosity\\ \cline{6-7} ID & Model & (\Msol)	& (\Msol) & ( \kmsec)
& (kpc) 	& (arcsec) 	 & 	($\times 10^{40} \ergs$)\\
\hline
1 & momentum (MDW) & 7.70$\times10^{11}$ & 2.03$\times10^{10}$ & 206.7 & 77.5 & 9.8 & 882.7\\
 & strong (SW) & 7.24$\times10^{11}$ & 1.38$\times10^{10}$ & 202.5 & 76.0 & 9.6 & 830.9\\
 & weak (WW) & 8.15$\times10^{11}$ & 2.80$\times10^{10}$ & 210.7 & 79.0 & 10.0 & 935.3\\
 & none (NW) & 8.12$\times10^{11}$ & 2.00$\times10^{10}$ & 210.4 & 78.9 & 10.0 & 931.5\\
2 & MDW & 1.54$\times10^{11}$ & 3.36$\times10^{9}$ & 120.9 & 45.3 & 5.8 & 176.5\\
 & SW & 1.40$\times10^{11}$ & 1.07$\times10^{9}$ & 117.2 & 43.9 & 5.6 & 160.8\\
 & WW & 1.70$\times10^{11}$ & 6.22$\times10^{9}$ & 125.0 & 46.9 & 5.9 & 195.1\\
 & NW & 1.65$\times10^{11}$ & 4.76$\times10^{9}$ & 123.7 & 46.4 & 5.9 & 189.3\\
3 & MDW & 1.46$\times10^{10}$ & 2.74$\times10^{8}$ & 55.1 & 20.7 & 2.6 & 16.7\\
 & SW & 1.36$\times10^{10}$ & 1.30$\times10^{8}$ & 53.8 & 20.2 & 2.6 & 15.6\\
 & WW & 1.58$\times10^{10}$ & 8.10$\times10^{8}$ & 56.6 & 21.2 & 2.7 & 18.2\\
 & NW & 1.57$\times10^{10}$ & 9.12$\times10^{8}$ & 56.5 & 21.2 & 2.7 & 18.1\\
\hline
\end{tabular}
\caption{Characteristic properties of three DM haloes of DLA host galaxies.}
\label{haloprop}
\end{table*}


\subsection{Three representative haloes} \label{S:3haloes} 
 
We have selected three haloes at $z = 3$ from the simulations for 
detailed study. Some basic properties of the haloes are summarised in 
Table \ref{haloprop}. The virial velocity and radius are calculated from 
the mass of the particles grouped by the halo finder --- see 
\citet{2004MNRAS.355..694M} for the relevant formulae.
Note that the properties of a given halo 
for the different wind model simulations are very similar. 
The final column lists the \lya luminosity ($L_{\lya}$), calculated following 
\citetalias{2009MNRAS.397..511B} and \citetalias{2010MNRAS.403..870B} as,
\begin{equation} L_{\lya} = 10^{42} \left( \frac{v_\ro{c}}{100
 \kmsec}\right)^3 ~ \ergs
\end{equation}
where $v_\ro{c} \propto M^{1/3}$ is the virial velocity
of the halo. With this choice \citetalias{2009MNRAS.397..511B} and 
\citetalias{2010MNRAS.403..870B} were able to reproduce the
luminosity function of the \citetalias{2008ApJ...681..856R} emitters. 
The masses of the three haloes probe the range of masses 
predicted to contribute significantly to the incidence rate of DLAs /
emitters per unit log mass ($\dd^2
\cN / \dd X / \dd \log_{10} M$). The corresponding luminosities span
the range of those observed in the survey of \citetalias{2008ApJ...681..856R}.

The physical properties of these haloes (neutral hydrogen
density, temperature and bulk velocity) were projected onto a
regular 3D grid centred on the centre of mass of the halo. 
The cubes containing Haloes 1 and 2 have $128^3$ cells, each with 
side length $3.125 h^{-1}$ comov.kpc, making 
the entire cube $400 h^{-1}$ comov.kpc across. The cube 
containing Halo 3 has $64^3$ cells, each also $3.125 h^{-1}$ comov.kpc 
across, making the with of the entire cube $200 h^{-1}$ comov.kpc.

\begin{figure*} \centering
\mnpg{
%
%
\begin{psfrags}%
\psfragscanon%
%
\psfrag{s05}[t][t]{\color[rgb]{0,0,0}\setlength{\tabcolsep}{0pt}\begin{tabular}{c}radius [arcsec]\end{tabular}}%
\psfrag{s06}[b][b]{\color[rgb]{0,0,0}\setlength{\tabcolsep}{0pt}\begin{tabular}{c}$dM / dt$   $[M_{sun} / yr]$\end{tabular}}%
\psfrag{s10}[][]{\color[rgb]{0,0,0}\setlength{\tabcolsep}{0pt}\begin{tabular}{c} \end{tabular}}%
\psfrag{s11}[][]{\color[rgb]{0,0,0}\setlength{\tabcolsep}{0pt}\begin{tabular}{c} \end{tabular}}%
\psfrag{s12}[l][l]{\color[rgb]{0,0,0}Halo 3}%
\psfrag{s13}[l][l]{\color[rgb]{0,0,0}Halo 1}%
\psfrag{s14}[l][l]{\color[rgb]{0,0,0}Halo 2}%
\psfrag{s15}[l][l]{\color[rgb]{0,0,0}Halo 3}%
%
\psfrag{x01}[t][t]{0}%
\psfrag{x02}[t][t]{2}%
\psfrag{x03}[t][t]{4}%
\psfrag{x04}[t][t]{6}%
\psfrag{x05}[t][t]{8}%
%
\psfrag{v01}[r][r]{-1}%
\psfrag{v02}[r][r]{-0.5}%
\psfrag{v03}[r][r]{0}%
\psfrag{v04}[r][r]{0.5}%
\psfrag{v05}[r][r]{1}%
%
\includegraphics[width=\textwidth]{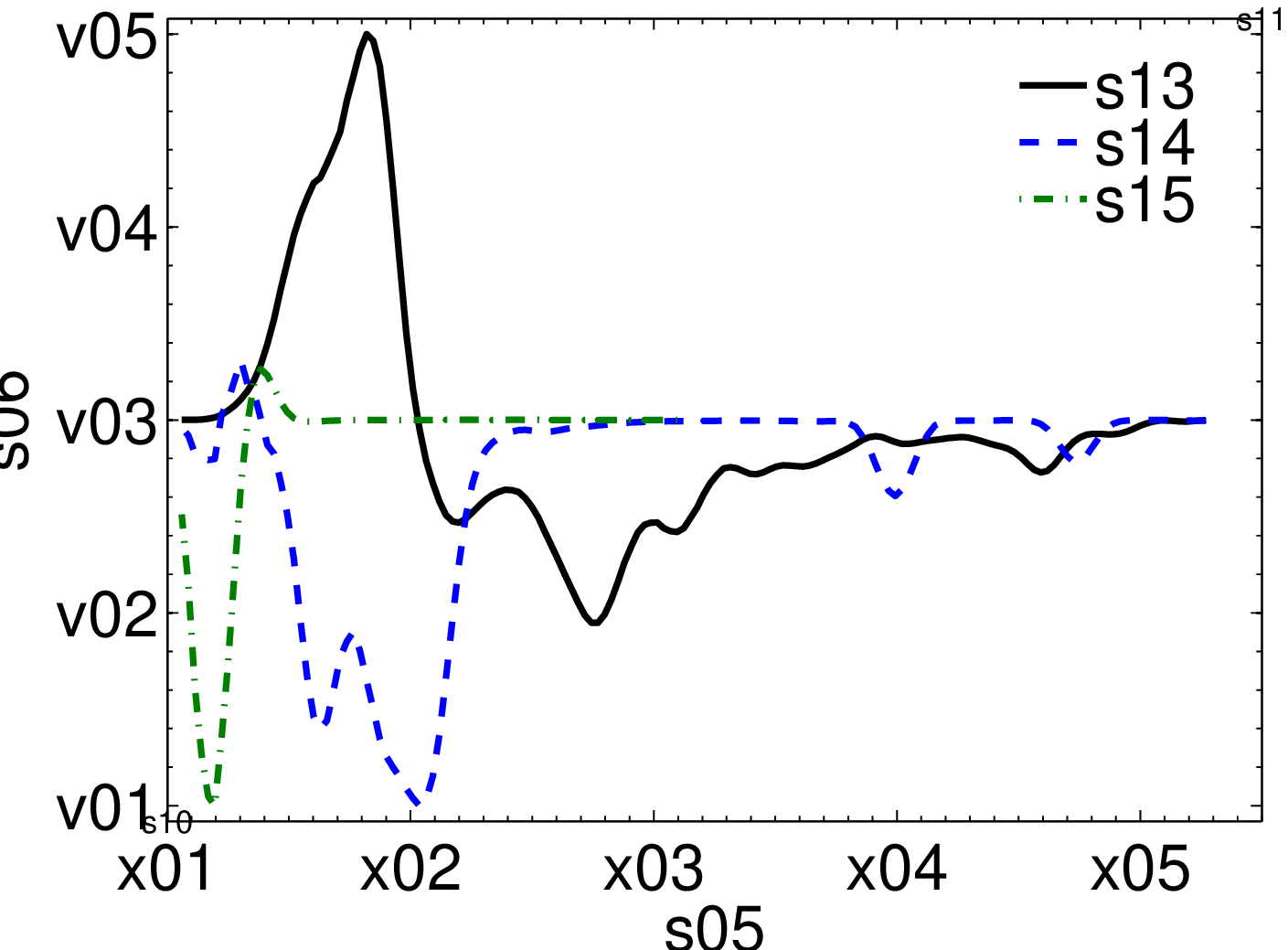}
\end{psfrags}%
%
} \hsp
\mnpg{
%
%
\begin{psfrags}%
\psfragscanon%
%
\psfrag{s03}[t][t]{\color[rgb]{0,0,0}\setlength{\tabcolsep}{0pt}\begin{tabular}{c}impact parameter [arcsec]\end{tabular}}%
\psfrag{s04}[b][b]{\color[rgb]{0,0,0}\setlength{\tabcolsep}{0pt}\begin{tabular}{c}log HI Column density [cm$^{-2}$]\end{tabular}}%
%
\psfrag{x01}[t][t]{0}%
\psfrag{x02}[t][t]{2}%
\psfrag{x03}[t][t]{4}%
\psfrag{x04}[t][t]{6}%
\psfrag{x05}[t][t]{8}%
%
\psfrag{v01}[r][r]{14}%
\psfrag{v02}[r][r]{16}%
\psfrag{v03}[r][r]{18}%
\psfrag{v04}[r][r]{20}%
%
\includegraphics[width=\textwidth]{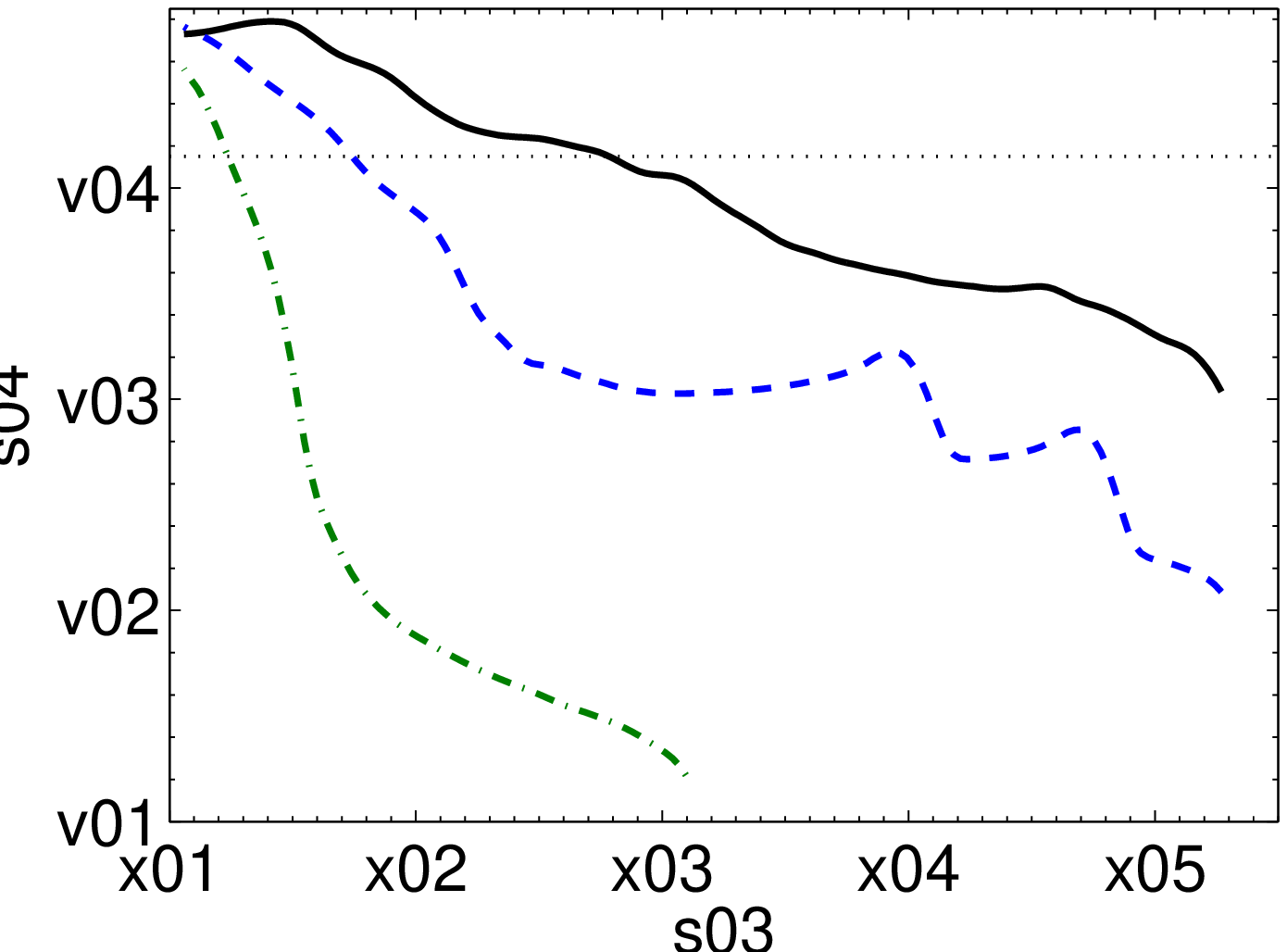}
\end{psfrags}%
%
} \vsp
\mnpg{
%
%
\begin{psfrags}%
\psfragscanon%
%
\psfrag{s03}[t][t]{\color[rgb]{0,0,0}\setlength{\tabcolsep}{0pt}\begin{tabular}{c}$v_{ph}$ [km/s]\end{tabular}}%
\psfrag{s04}[b][b]{\color[rgb]{0,0,0}\setlength{\tabcolsep}{0pt}\begin{tabular}{c}P$_{v} \times 10^3$ [km/s]$^{-1}$\end{tabular}}%
%
\psfrag{x01}[t][t]{-1000}%
\psfrag{x02}[t][t]{-500}%
\psfrag{x03}[t][t]{0}%
\psfrag{x04}[t][t]{500}%
\psfrag{x05}[t][t]{1000}%
%
\psfrag{v01}[r][r]{0}%
\psfrag{v02}[r][r]{0.5}%
\psfrag{v03}[r][r]{1}%
\psfrag{v04}[r][r]{1.5}%
\psfrag{v05}[r][r]{2}%
%
\includegraphics[width=\textwidth]{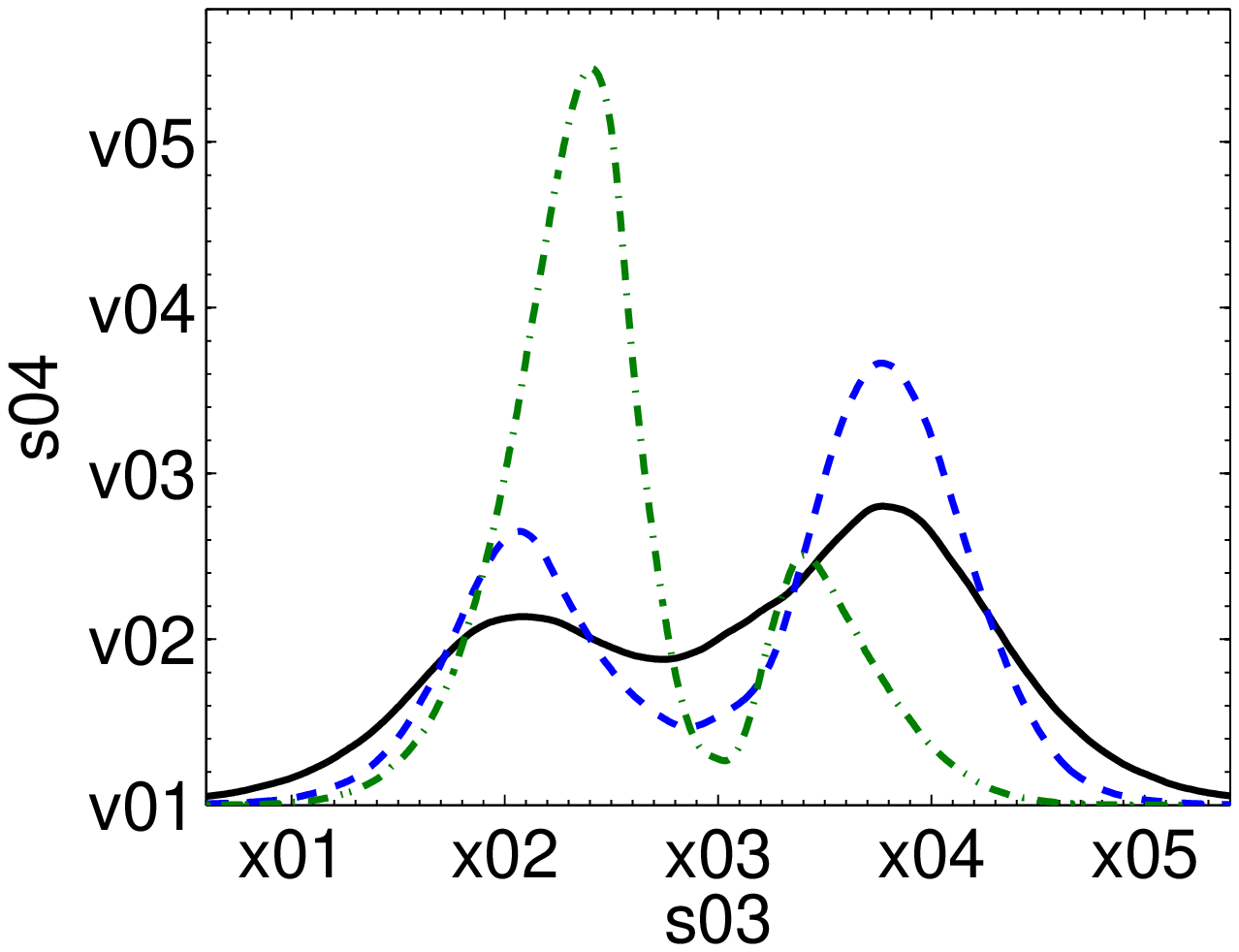}
\end{psfrags}%
%
} \hsp
\mnpg{
%
%
\begin{psfrags}%
\psfragscanon%
%
\psfrag{s03}[t][t]{\color[rgb]{0,0,0}\setlength{\tabcolsep}{0pt}\begin{tabular}{c}impact parameter [arcsec]\end{tabular}}%
\psfrag{s04}[b][b]{\color[rgb]{0,0,0}\setlength{\tabcolsep}{0pt}\begin{tabular}{c}log S   [erg/s/cm$^2$/arcsec$^2$]\end{tabular}}%
%
\psfrag{x01}[t][t]{0}%
\psfrag{x02}[t][t]{2}%
\psfrag{x03}[t][t]{4}%
\psfrag{x04}[t][t]{6}%
\psfrag{x05}[t][t]{8}%
%
\psfrag{v01}[r][r]{-22}%
\psfrag{v02}[r][r]{-20}%
\psfrag{v03}[r][r]{-18}%
\psfrag{v04}[r][r]{-16}%
%
\includegraphics[width=\textwidth]{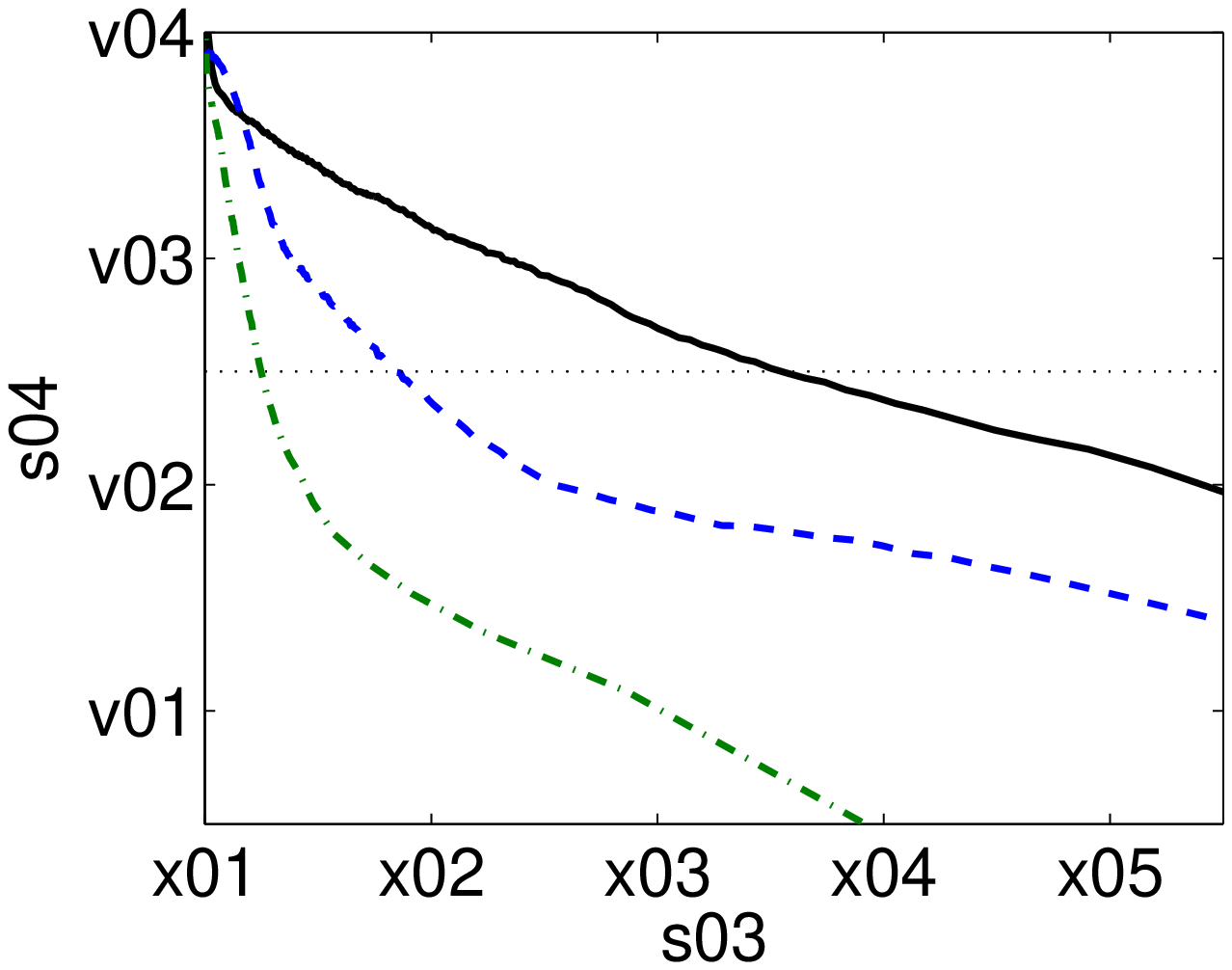}
\end{psfrags}%
%
}
	\caption[Angularly-averaged properties of each of the three
haloes]{The angularly-averaged properties of the three DLA host
 galaxies described in Table 1, for the MDW wind model. The
upper left panel shows $\dd M/ \dd t$, which is the rate at which \hi mass
is flowing through a surface of constant radius $r$, normalised to a
maximum $|\dot{M}| = 1$. The maximum $\dot{M}$ (by which the 
curve are scaled) are 97, 8.5 and 1.5 $\Msol /$yr, for Halo 1, 2 and 3
respectively.
The upper right panel shows the average column density
for a sightline passing at a given impact parameter from the centre of
the halo. The thin dotted horizontal line indicates the minimum column
density of a DLA, $N_\ro{DLA} = 10^{20.3} \cm$. The lower left panel
shows the (observed) angularly-averaged spectrum, with the area under
each curve normalised to unity. The lower right panel shows the
surface brightness profile; the horizontal line shows the surface 
brightness limit of the \citetalias{2008ApJ...681..856R} survey. The assumed total 
\lya luminosities (which normalise each curve) are given in the last
column of Table \ref{haloprop}.} \label{fig:allplots}
\end{figure*}

We will first examine the simulation with the MDW implementation of galactic winds,
before comparing with the other simulations in later sections. 
Figure \ref{fig:allplots} shows the angularly-averaged radial mass flow
rate, projected column density of neutral hydrogen and the 
\lya surface brightness as a function of radius/impact parameter as
well as the angularly-averaged spectral shape of the \lya
emission for the three halos. 

The rate at which \hi mass flows through a surface of constant radius
$r$ is shown in the upper left panel of Figure \ref{fig:allplots}, 
normalised to a maximum $|\dot{M}| = 1$. Positive (negative) values correspond to
outflow (inflow). The maximum $\dot{M}$ is
97, 8.5 and 1.5 $\Msol /$yr, for Halo 1, 2 and 3 respectively. 
Note that, for the MDW implementation shown here, only in the most massive halo
does the wind actually manage to globally reverse the gravitational
infall. The mass flow shows a mixture of inflow and outflow, with inflow often
dominating in the outer parts of the halo. 


The upper right panel of Figure \ref{fig:allplots}
shows the average column density for a sightline
passing at a given impact parameter from the centre of the halo. The
thin dotted horizontal line indicates the minimum column density of a
DLA, $N_\ro{DLA} = 10^{20.3} \cm$. As expected, the column density
peaks near the centre of the halo. The
``steps'' in the column density profiles of Haloes 2 and 3 show the
influence of separate clumps of \hi at large radii.
The neutral hydrogen column density exceeds the threshold for a DLA,
which extends to a radius of 3.5, 1.5, and 0.5 arcsec, respectively. 

The lower left panel shows the
(observed) angularly-averaged spectrum, with each curve normalised to
unity. The lower right panel shows the surface brightness profile. The
luminosity used to normalise each curve is given in the last column of
Table \ref{haloprop}. 

The surface brightness profiles peak at the centre of the halo, and generally
follow the decline of the column density whilst being much
smoother. 
The \lya emission is scattered to similarly large radii as in the
observed emitters. The haloes are surrounded by low surface brightness emission 
which extends to radii of several arcsec. 
The size of the emitter, as measured to the surface brightness limit of the
\citetalias{2008ApJ...681..856R} (shown by the horizontal dotted line), 
is generally larger than the cross-section for damped
absorption. The more massive haloes are more spatially extended, while
the smaller haloes display a more pronounced central peak.

The angularly-averaged spectrum of the most massive halo is 
fairly symmetric with well separated blue and red peaks. 
The red peak is slightly stronger. 
The two low mass haloes have a 
more clearly dominant red and blue peak, respectively. Increasing
central column density clearly leads to more 
diffusion in frequency space. 
We will see in the next section that the spectral shape 
varies strongly between different directions and 
that there is a correlation between the dominance of inflow (outflow) 
and the domination of the blue (red) peak in the spectrum.
We will leave a more detailed discussion until then. 


\section{The effect of galactic winds on \lya emission} \label{S:galwind} 

We will now discuss the simulations of the intermediate mass halo (our fiducial Halo 2) in 
more detail with particular emphasis on the effect of
galactic winds on the physical properties of the gas in the halo 
and on the emission properties of the \lya emission. 
Halo 2 has a mass of about $1.5 \times 10^{10} \Msol$ and a virial velocity of
about $120 \kmsec$. Its properties are similar to the brightest
of the emitters in the \citetalias{2008ApJ...681..856R} survey, and its mass coincides with the
peak of our preferred model for the DLA mass function $(\dd^2 \cN /
\dd X / \dd \log_{10} M)$ from \citetalias{2010MNRAS.403..870B}
and thus should represent a typical DLA host galaxy. 


\subsection{Angularly-Averaged Properties} \label{S:angav} 

\begin{figure*} \centering
		\mnpg{
%
%
\begin{psfrags}%
\psfragscanon%
%
\psfrag{s03}[t][t]{\color[rgb]{0,0,0}\setlength{\tabcolsep}{0pt}\begin{tabular}{c}radius [arcsec]\end{tabular}}%
\psfrag{s04}[b][b]{\color[rgb]{0,0,0}\setlength{\tabcolsep}{0pt}\begin{tabular}{c}log HI number density [cm$^{-3}$]\end{tabular}}%
%
\psfrag{x01}[t][t]{0}%
\psfrag{x02}[t][t]{2}%
\psfrag{x03}[t][t]{4}%
\psfrag{x04}[t][t]{6}%
\psfrag{x05}[t][t]{8}%
%
\psfrag{v01}[r][r]{-6}%
\psfrag{v02}[r][r]{-4}%
\psfrag{v03}[r][r]{-2}%
\psfrag{v04}[r][r]{0}%
%
\includegraphics[width=\textwidth]{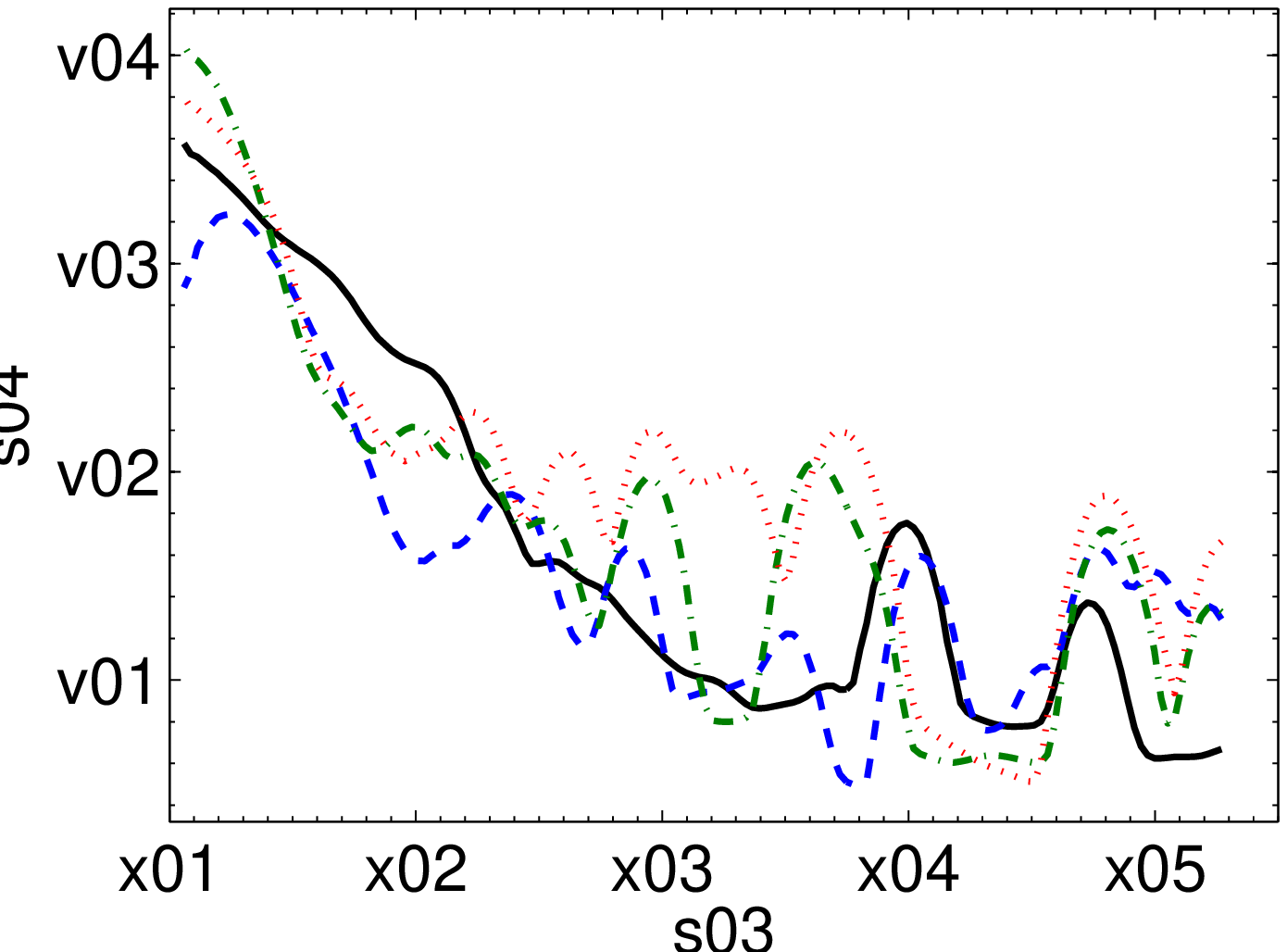}
\end{psfrags}%
%
}\hsp
		\mnpg{
%
%
\begin{psfrags}%
\psfragscanon%
%
\psfrag{s05}[t][t]{\color[rgb]{0,0,0}\setlength{\tabcolsep}{0pt}\begin{tabular}{c}impact parameter [arcsec]\end{tabular}}%
\psfrag{s06}[b][b]{\color[rgb]{0,0,0}\setlength{\tabcolsep}{0pt}\begin{tabular}{c}log HI Column density [cm$^{-2}$]\end{tabular}}%
\psfrag{s10}[][]{\color[rgb]{0,0,0}\setlength{\tabcolsep}{0pt}\begin{tabular}{c} \end{tabular}}%
\psfrag{s11}[][]{\color[rgb]{0,0,0}\setlength{\tabcolsep}{0pt}\begin{tabular}{c} \end{tabular}}%
\psfrag{s12}[l][l]{\color[rgb]{0,0,0}NW}%
\psfrag{s13}[l][l]{\color[rgb]{0,0,0}MDW}%
\psfrag{s14}[l][l]{\color[rgb]{0,0,0}SW}%
\psfrag{s15}[l][l]{\color[rgb]{0,0,0}WW}%
\psfrag{s16}[l][l]{\color[rgb]{0,0,0}NW}%
%
\psfrag{x01}[t][t]{0}%
\psfrag{x02}[t][t]{2}%
\psfrag{x03}[t][t]{4}%
\psfrag{x04}[t][t]{6}%
\psfrag{x05}[t][t]{8}%
%
\psfrag{v01}[r][r]{16}%
\psfrag{v02}[r][r]{17}%
\psfrag{v03}[r][r]{18}%
\psfrag{v04}[r][r]{19}%
\psfrag{v05}[r][r]{20}%
\psfrag{v06}[r][r]{21}%
\psfrag{v07}[r][r]{22}%
%
\includegraphics[width=\textwidth]{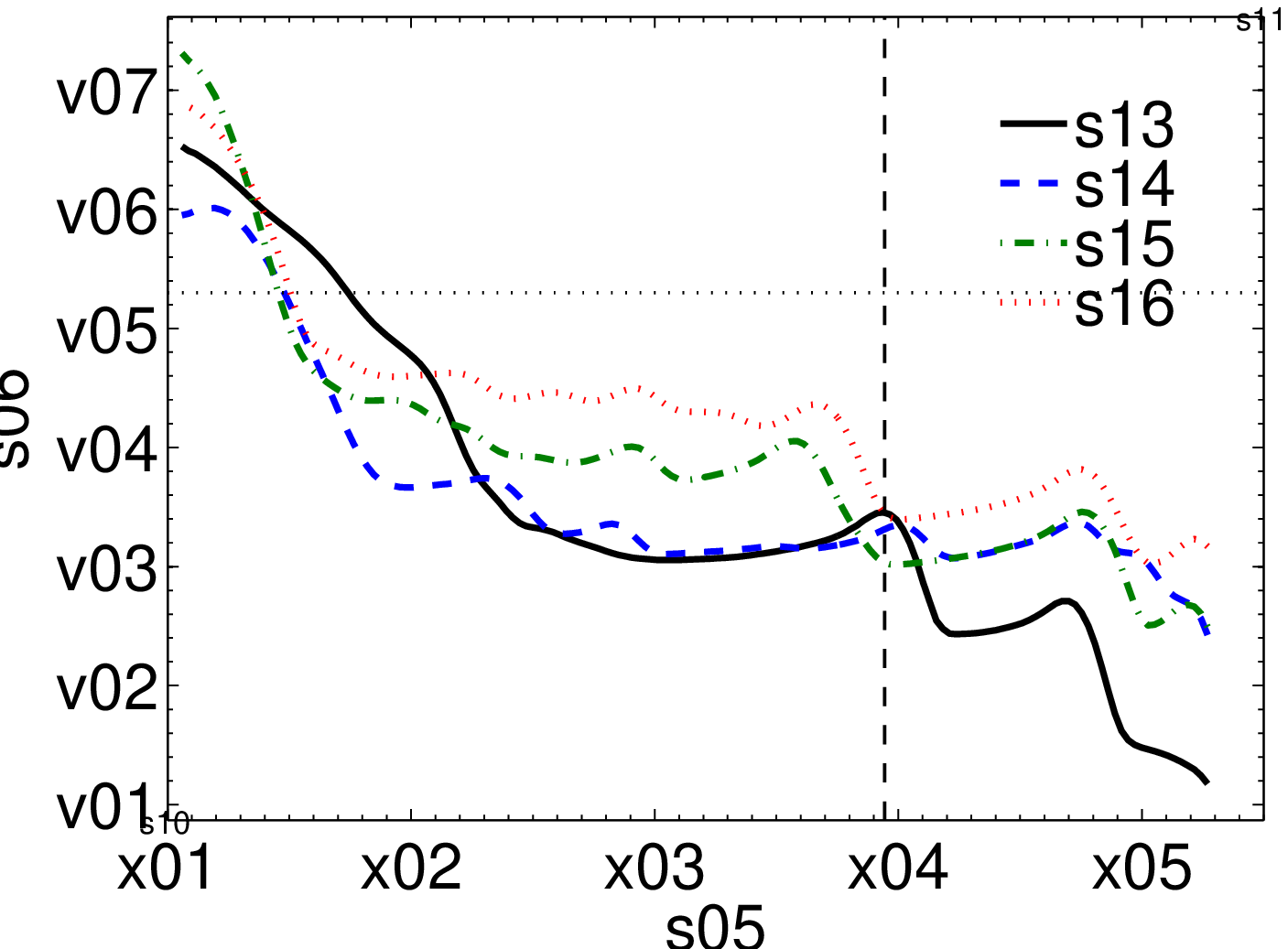}
\end{psfrags}%
%
}
		\mnpg{
%
%
\begin{psfrags}%
\psfragscanon%
%
\psfrag{s03}[t][t]{\color[rgb]{0,0,0}\setlength{\tabcolsep}{0pt}\begin{tabular}{c}radius [arcsec]\end{tabular}}%
\psfrag{s04}[b][b]{\color[rgb]{0,0,0}\setlength{\tabcolsep}{0pt}\begin{tabular}{c}radial velocity (dens. wgt.) [km/s]\end{tabular}}%
%
\psfrag{x01}[t][t]{0}%
\psfrag{x02}[t][t]{2}%
\psfrag{x03}[t][t]{4}%
\psfrag{x04}[t][t]{6}%
\psfrag{x05}[t][t]{8}%
%
\psfrag{v01}[r][r]{-150}%
\psfrag{v02}[r][r]{-100}%
\psfrag{v03}[r][r]{-50}%
\psfrag{v04}[r][r]{0}%
\psfrag{v05}[r][r]{50}%
\psfrag{v06}[r][r]{100}%
%
\includegraphics[width=\textwidth]{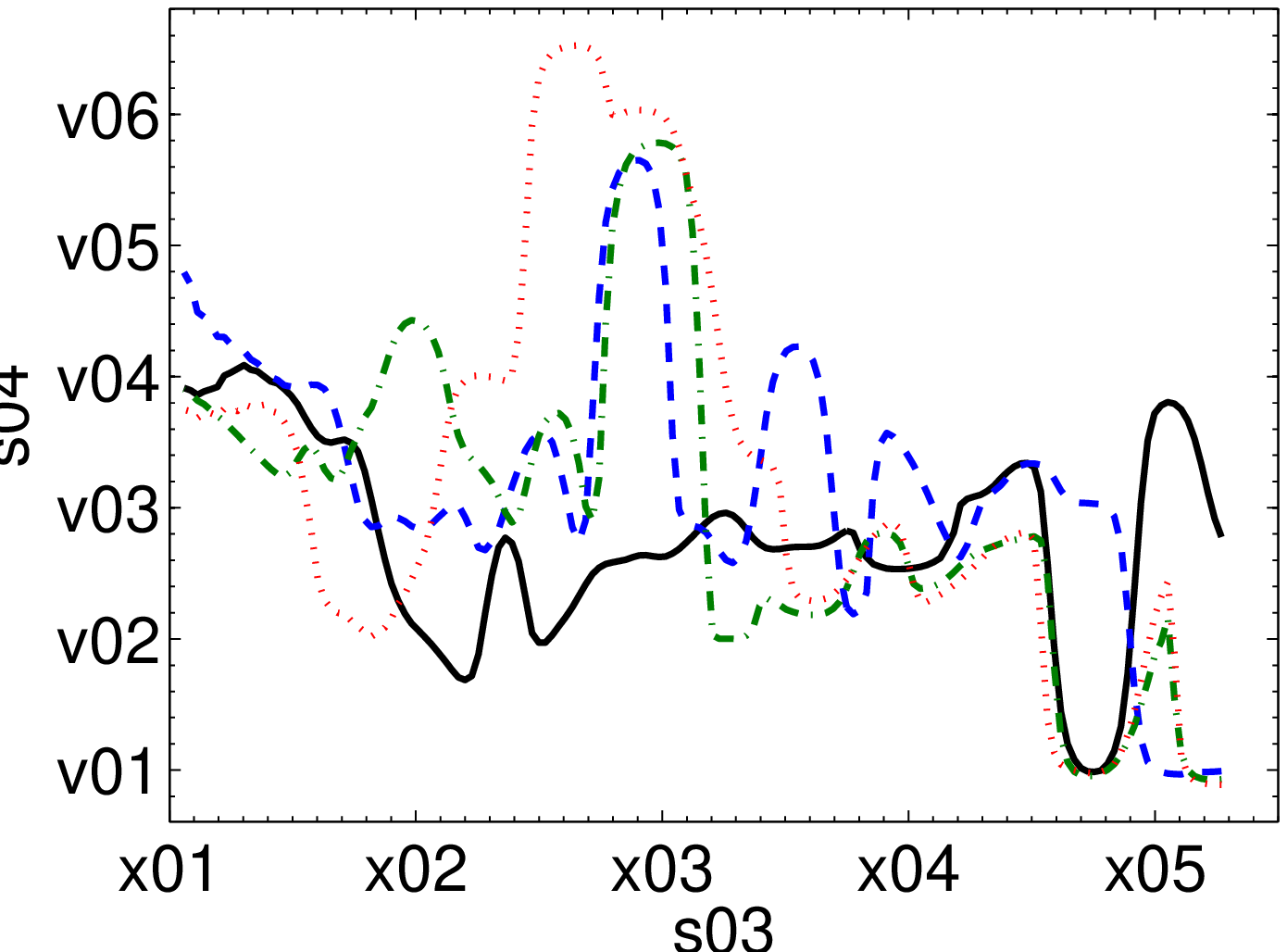}
\end{psfrags}%
%
}\hsp
		\mnpg{
%
%
\begin{psfrags}%
\psfragscanon%
%
\psfrag{s03}[t][t]{\color[rgb]{0,0,0}\setlength{\tabcolsep}{0pt}\begin{tabular}{c}radius [arcsec]\end{tabular}}%
\psfrag{s04}[b][b]{\color[rgb]{0,0,0}\setlength{\tabcolsep}{0pt}\begin{tabular}{c}radial velocity (vol. wgt.) [km/s]\end{tabular}}%
%
\psfrag{x01}[t][t]{0}%
\psfrag{x02}[t][t]{2}%
\psfrag{x03}[t][t]{4}%
\psfrag{x04}[t][t]{6}%
\psfrag{x05}[t][t]{8}%
%
\psfrag{v01}[r][r]{-50}%
\psfrag{v02}[r][r]{0}%
\psfrag{v03}[r][r]{50}%
\psfrag{v04}[r][r]{100}%
\psfrag{v05}[r][r]{150}%
\psfrag{v06}[r][r]{200}%
\psfrag{v07}[r][r]{250}%
%
\includegraphics[width=\textwidth]{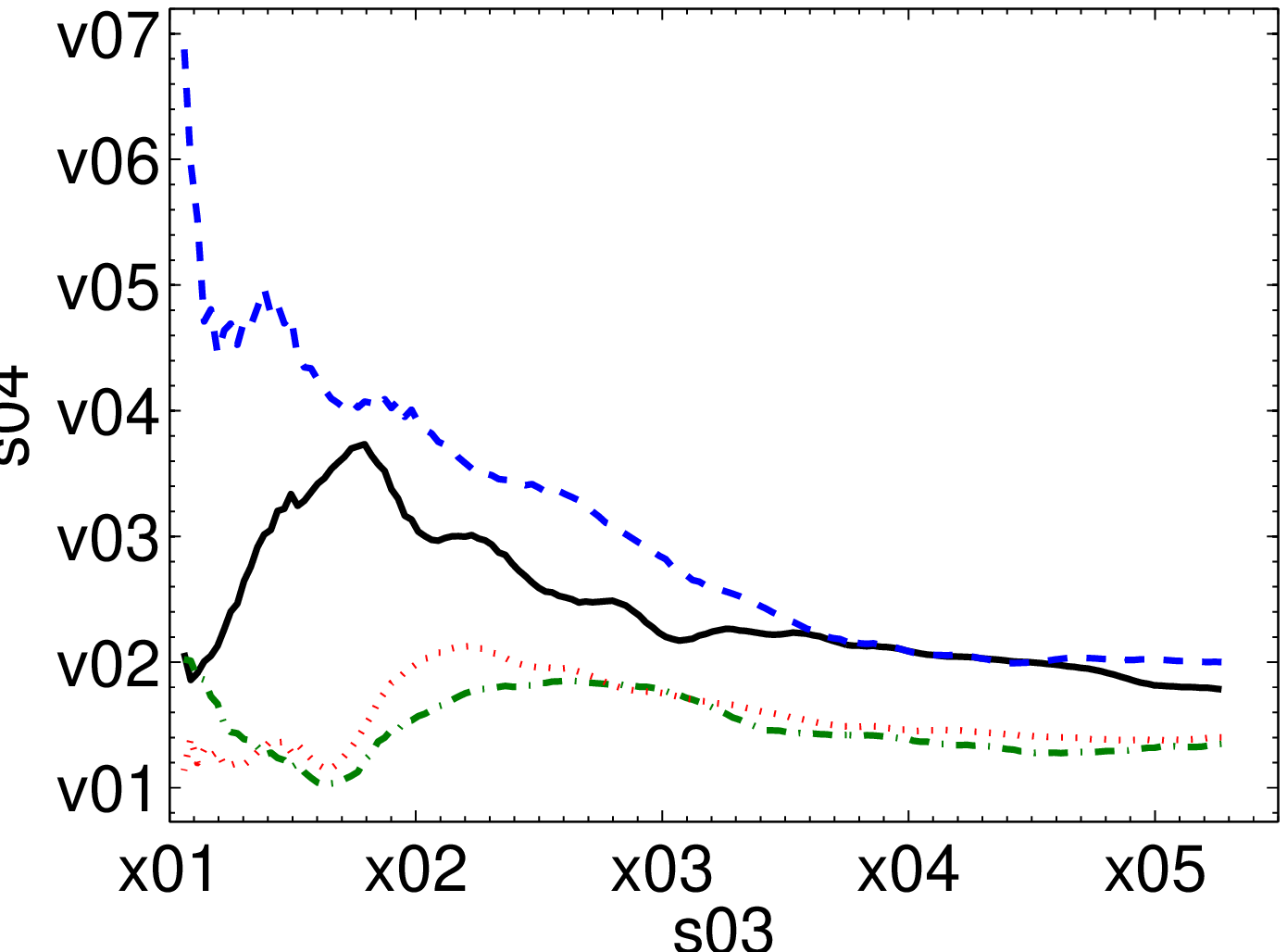}
\end{psfrags}%
%
}
		\mnpg{
%
%
\begin{psfrags}%
\psfragscanon%
%
\psfrag{s03}[t][t]{\color[rgb]{0,0,0}\setlength{\tabcolsep}{0pt}\begin{tabular}{c}radius [arcsec]\end{tabular}}%
\psfrag{s04}[b][b]{\color[rgb]{0,0,0}\setlength{\tabcolsep}{0pt}\begin{tabular}{c}$dM / dt$   $[M_{sun} / yr]$\end{tabular}}%
%
\psfrag{x01}[t][t]{0}%
\psfrag{x02}[t][t]{2}%
\psfrag{x03}[t][t]{4}%
\psfrag{x04}[t][t]{6}%
\psfrag{x05}[t][t]{8}%
%
\psfrag{v01}[r][r]{-25}%
\psfrag{v02}[r][r]{-20}%
\psfrag{v03}[r][r]{-15}%
\psfrag{v04}[r][r]{-10}%
\psfrag{v05}[r][r]{-5}%
\psfrag{v06}[r][r]{0}%
\psfrag{v07}[r][r]{5}%
%
\includegraphics[width=\textwidth]{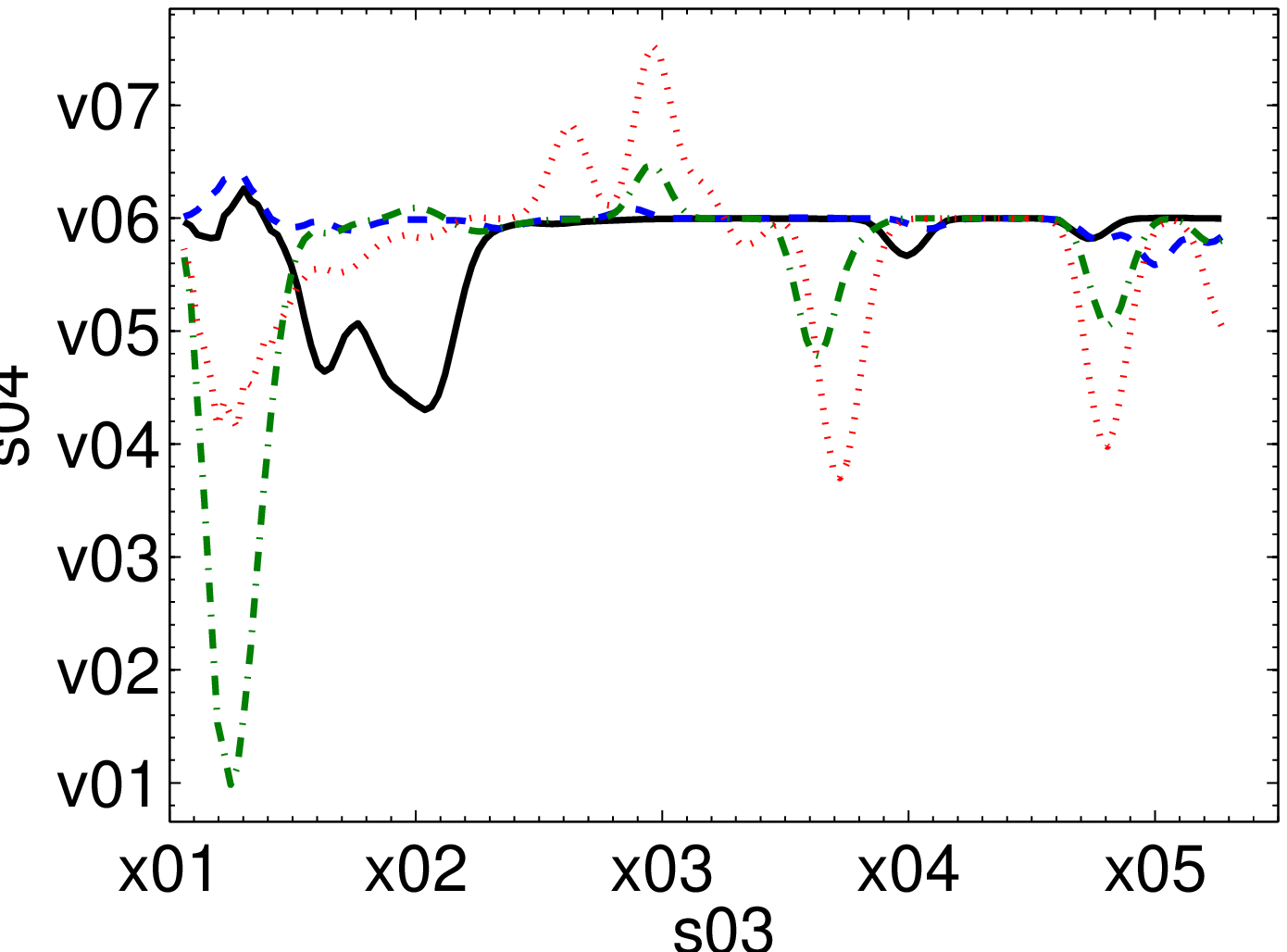}
\end{psfrags}%
%
}\hsp
		\mnpg{
%
%
\begin{psfrags}%
\psfragscanon%
%
\psfrag{s03}[t][t]{\color[rgb]{0,0,0}\setlength{\tabcolsep}{0pt}\begin{tabular}{c}radius [arcsec]\end{tabular}}%
\psfrag{s04}[b][b]{\color[rgb]{0,0,0}\setlength{\tabcolsep}{0pt}\begin{tabular}{c}rotation velocity [km/s]\end{tabular}}%
%
\psfrag{x01}[t][t]{0}%
\psfrag{x02}[t][t]{2}%
\psfrag{x03}[t][t]{4}%
\psfrag{x04}[t][t]{6}%
\psfrag{x05}[t][t]{8}%
%
\psfrag{v01}[r][r]{0}%
\psfrag{v02}[r][r]{50}%
\psfrag{v03}[r][r]{100}%
\psfrag{v04}[r][r]{150}%
\psfrag{v05}[r][r]{200}%
%
\includegraphics[width=\textwidth]{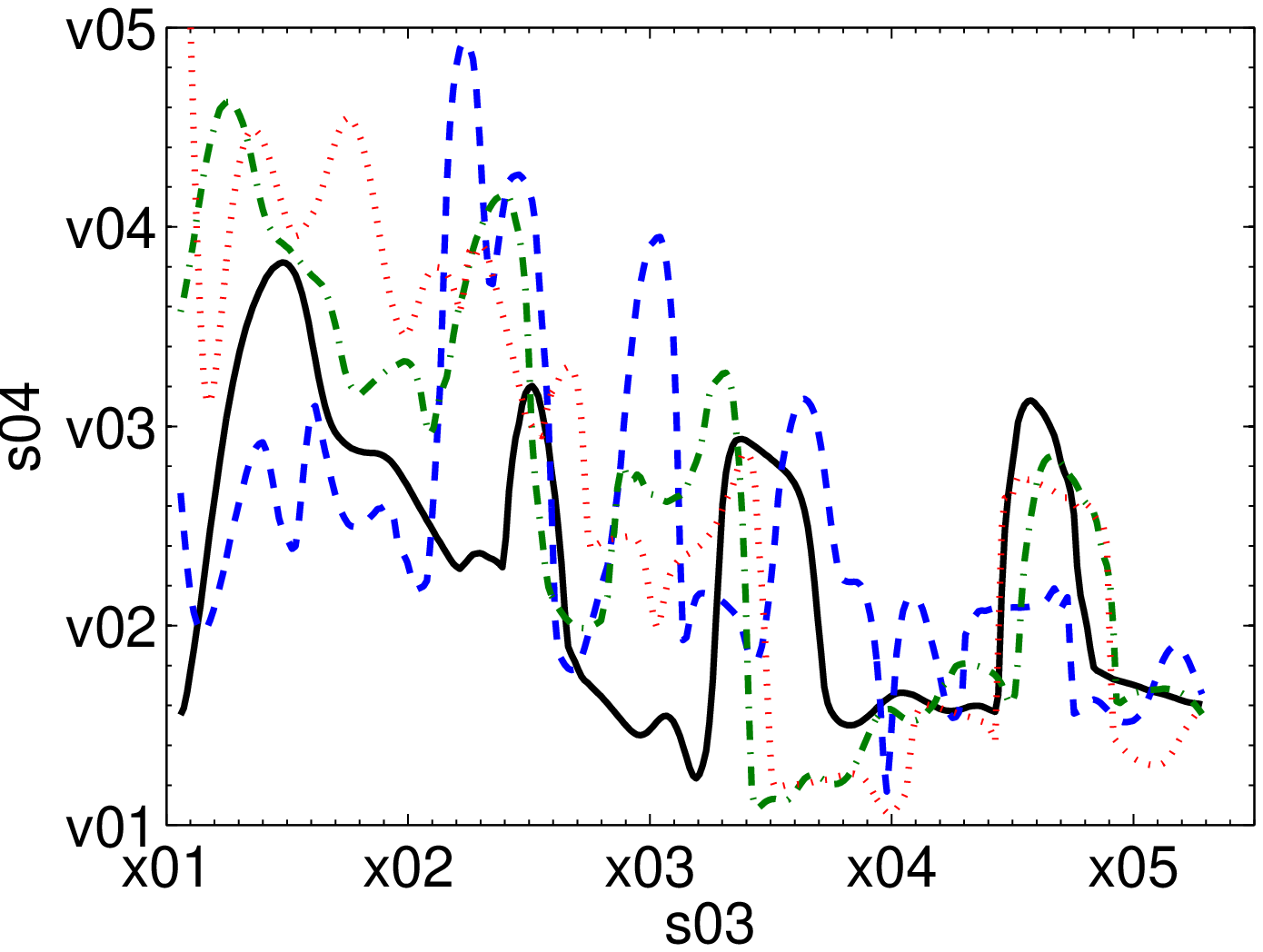}
\end{psfrags}%
%
}
		\mnpg{
%
%
\begin{psfrags}%
\psfragscanon%
%
\psfrag{s03}[t][t]{\color[rgb]{0,0,0}\setlength{\tabcolsep}{0pt}\begin{tabular}{c}$v_{ph}$ [km/s]\end{tabular}}%
\psfrag{s04}[b][b]{\color[rgb]{0,0,0}\setlength{\tabcolsep}{0pt}\begin{tabular}{c}P$_{v} \times 10^3$ [km/s]$^{-1}$\end{tabular}}%
%
\psfrag{x01}[t][t]{-1000}%
\psfrag{x02}[t][t]{0}%
\psfrag{x03}[t][t]{1000}%
%
\psfrag{v01}[r][r]{0}%
\psfrag{v02}[r][r]{0.5}%
\psfrag{v03}[r][r]{1}%
\psfrag{v04}[r][r]{1.5}%
\psfrag{v05}[r][r]{2}%
%
\includegraphics[width=\textwidth]{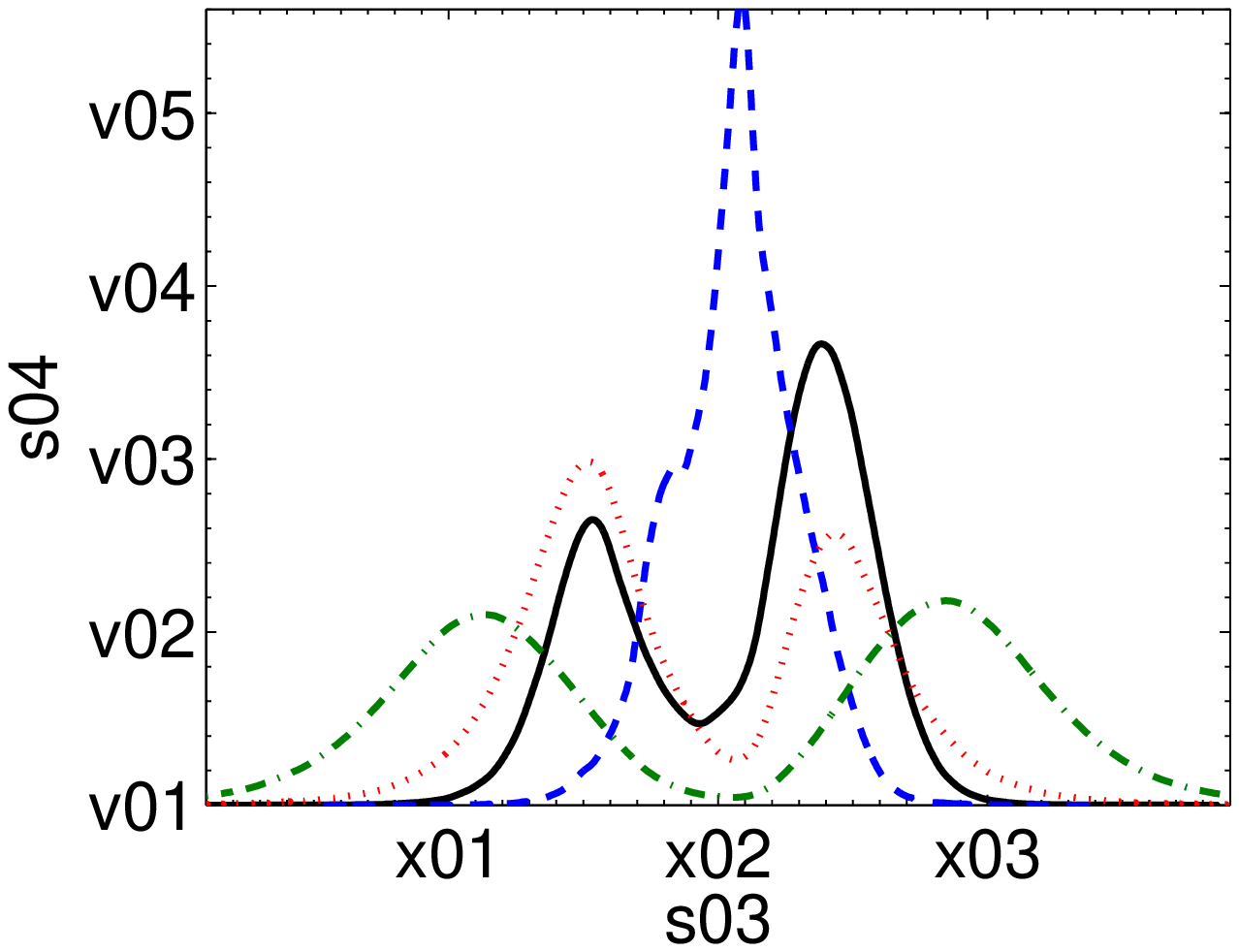}
\end{psfrags}%
%
}\hsp
		\mnpg{
%
%
\begin{psfrags}%
\psfragscanon%
%
\psfrag{s03}[t][t]{\color[rgb]{0,0,0}\setlength{\tabcolsep}{0pt}\begin{tabular}{c}impact parameter [arcsec]\end{tabular}}%
\psfrag{s04}[b][b]{\color[rgb]{0,0,0}\setlength{\tabcolsep}{0pt}\begin{tabular}{c}log S   [erg/s/cm$^2$/arcsec$^2$]\end{tabular}}%
%
\psfrag{x01}[t][t]{0}%
\psfrag{x02}[t][t]{2}%
\psfrag{x03}[t][t]{4}%
\psfrag{x04}[t][t]{6}%
\psfrag{x05}[t][t]{8}%
%
\psfrag{v01}[r][r]{-22}%
\psfrag{v02}[r][r]{-20}%
\psfrag{v03}[r][r]{-18}%
\psfrag{v04}[r][r]{-16}%
%
\includegraphics[width=\textwidth]{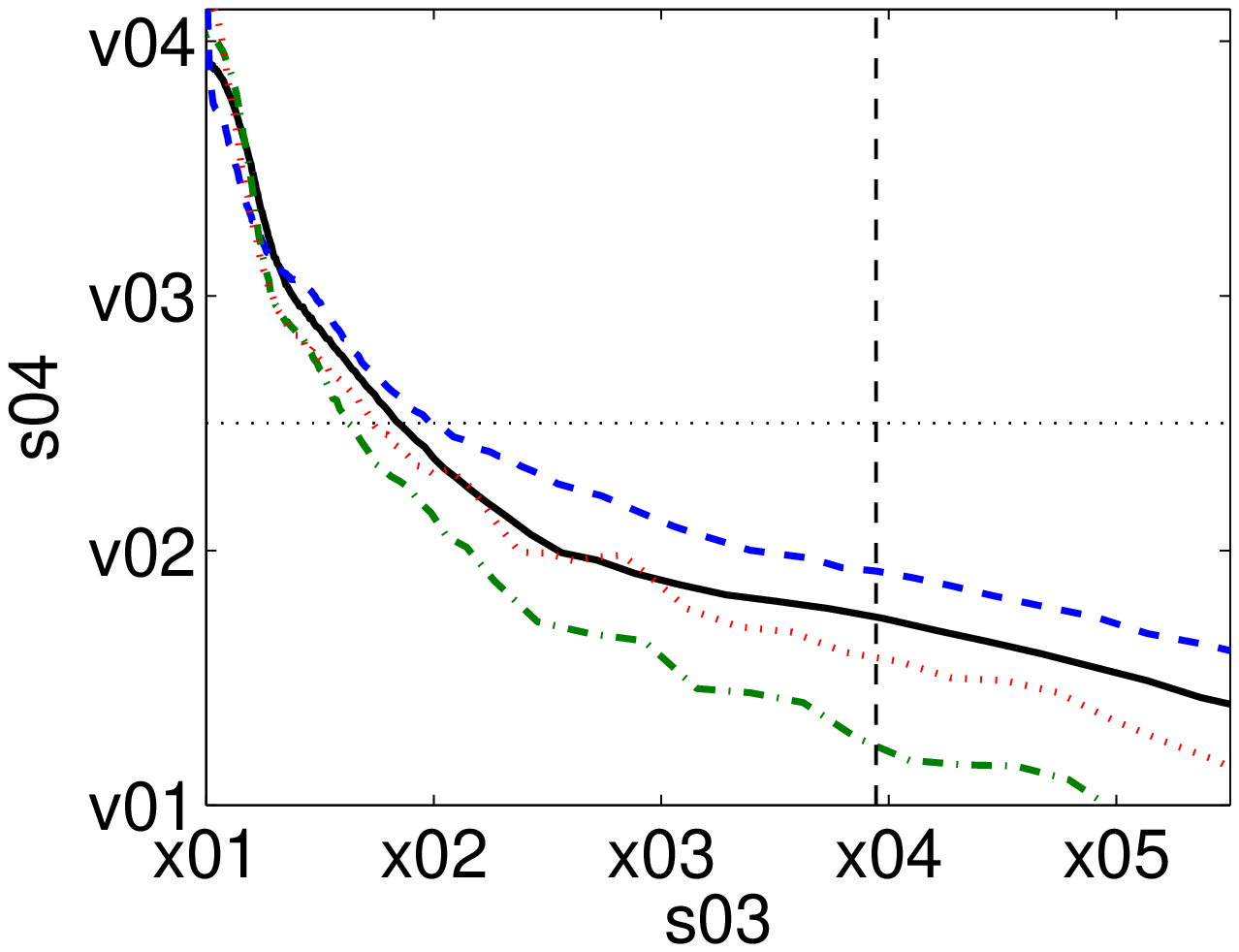}
\end{psfrags}%
%
}
		\caption{Angularly-averaged properties of Halo 2, for the wind
prescriptions as shown in the legend (MDW: momentum driven
wind, SW: strong wind, WW: weak wind, NW: no wind). The velocities and
$dM/dt$ are all for \hi. Vertical dashed lines 
indicate the virial radius. For a detailed 
discussion, see Section \ref{S:angav}.} \label{fig:av2sw}
	\end{figure*}

Angularly-averaged properties of simulation based on Halo 2 for each of the wind models
are shown in Figure \ref{fig:av2sw}. The legend in the top right panel
shows the line-styles of the different wind models


The top left panel shows the \hi number density as
function of radius. Note that the average cosmic number density of hydrogen 
at $z \approx 3$ is $1.2 \times 10^{-5}\ro{ cm}^{-3}$, while the
volume-averaged neutral fraction of hydrogen in the IGM is $\lesssim
10^{-5}$. A small decrease in the number density at small radii is the
result of the centre of mass of the neutral hydrogen being offset from
the centre of mass of the halo. The number density profile shows the
effect of several clumps of \hi. The most dramatic difference
between the models is in the centre of the haloes, where the presence
of a strong wind can reduce the gas density substantially. 

The top right panel shows the averaged column density for a sightline
passing at a given impact parameter from the centre of the halo. The
thin dotted horizontal line indicates the minimum column density of a
DLA. The column density is a decreasing function of radius, with sightlines
containing DLAs probing the centre of the halo. The ``clumps''
noticeable in the number density distribution, cause steps in the 
angle-averaged column density distribution and contribute to 
the Lyman-Limit System cross-section. 

The next panel (row 2, column 1) shows the (\hi) density-weighted radial
velocity profile,
\begin{equation}
v_\hi(r) \equiv \langle \vv_\ro{b} \cdot \hat{\rv} \rangle \equiv 
\frac{\int_0^{2\pi} \int_0^{\pi} ~ (\vv_\ro{b} \cdot
\hat{\rv}) ~ \rho_{\hi}(r,\theta,\phi) \sin \theta \df \theta \df
\phi} {\int_0^{2\pi} \int_0^{\pi} ~ \rho_{\hi}(r,\theta,\phi) \sin
\theta \df \theta \df \phi} .
\end{equation}
The plot reveals the presence of a mixture of inflow ($v_r$ negative) and
outflow, with the innermost parts of the halo tending to show outflow
for the stronger winds (SW and MDW), and inflow for the weaker winds
(WW and NW). 
 
The typical velocities of inflow are 50 - 100 \kmsec,
with the largest velocities being comparable to the virial
velocity. There is a general trend toward lower velocities in the
centre of the halo. The velocity profile is, however, far from smooth.
The total velocity of the gas ($v =
\sqrt{ \langle |\vv_\ro{b}|^2 \rangle}$), by comparison, varies between 
100-150 \kmsec for the stronger winds, and 140 - 200 \kmsec for the 
weaker winds, with a very slight preference for larger velocities at the
centre of the halo.

The panel to the right shows the volume-weighted radial velocity profile 
($v_V$). For the diffusion of \lya photons, volume-weighted
velocities are actually more relevant than density/mass-weighted 
velocities as the photons tend to find low-density paths of escape, 
and so the velocity in the underdense regions is as important or
more important as that in high-density regions. We see a clear
difference in $v_V$ between the stronger and weaker 
wind models. The strong wind models have outflows that dominate the kinematics 
of the gas along most of the escape paths of the photons. For weak wind models, 
inflow dominates.
The next panel (row 3, column 1) shows $\dd M/ \dd t$ for \hi, which is related to the
density-weighted radial velocity as $\dd M/ \dd t = 4 \pi r^2 v_\hi(r)
\rho^{\ro{av}}_{\hi}(r)$, where $\rho^{\ro{av}}_{\hi}$
is the average
neutral hydrogen density at radius $r$. Notice that the flows in
the innermost parts of the halo, while being of modest velocity, carry
the most significant amounts of mass.

Comparison of $v_\hi(r)$, $v_V$ and $\dd M/ \dd t$ is quite instructive. The 
relative smoothness of $v_V$ indicates that a small number of dense clumps 
dominate the mass flow, moving through a relatively smooth background whose 
kinematics are shaped by the winds. For example, the MDW model (solid black) 
shows a $\sim 100$ \kmsec inflowing clump at 2 arcsec that carries a significant 
amount of mass. However, $v_V$ is still positive, showing that the wind is filling 
the remaining space and outflowing regardless. \lya photons escaping 
along low column density paths will generally not ``see'' such dense clumps. 

The panel to the right shows the rotation velocity of the gas, averaged over 
\emph{cylindrical} shells, which is approximated 
as follows. If the gas in the halo were in a rotating disk with a rotation 
velocity depending only on the cylindrical radius, then the angular momentum 
within a cylindrical shell $R, R + \dd R$ would be,
\begin{equation}
|\mathbf{L_\ro{rot}}| = v_{\ro{rot}}(R) ~ R \sum_i m_i ~ ,
\end{equation}
where the sum is over the cells inside the shell. Thus for each $R$, we divide the 
angular momentum by $R$ times the mass in the shell. As with $v_\hi(r)$, this 
distribution is far from smooth due to  the influence of dense clumps of gas. 
The velocity drops from $\sim 150$ \kmsec to $\sim 50$ \kmsec as we move outwards. 
The velocity is generally higher in the weaker wind models, possibly indicating 
that the stronger winds disrupt or delay the settling of gas into a disk. We can also 
calculate the fraction of the kinetic energy at each radius which is in the form of 
rotational kinetic energy. This quantity peaks at $\sim 0.7$ at the centre of the
halo, decreasing to $\sim 0.2$ in the outer parts. This shows that the approximations 
we made in calculating $v_{\ro{rot}}$ are most accurate at the centre of the halo, 
as expected.

The bottom left panel shows the angularly-averaged spectrum.
of the \lya emission. The thick
black curve shows the spectrum as a function of the rest-frame 
velocity offset ($v_{\ro{ph}} \equiv c \Delta \lambda / \lambda_0$). 
All curves are normalised to have unit area.
The weaker wind models show a classic
double-peaked profile (\citealt{1973MNRAS.162...43H,1990ApJ...350..216N}; 
see also \citealt{1981ApJ...244..406U}),
while the stronger wind models show a more
prominent red peak, indicating the influence of the outflowing gas.

We also see that the stronger galactic winds --- somewhat counterintuitively --- 
result in a narrower spectral distribution. This indicates that it is the 
high central column density, more than the higher outer gas velocities, that 
are the dominant factor in establishing the typical escape frequency of 
\lya photons.

The thick black curve in the bottom right panel shows the
angularly-averaged surface brightness profile. The luminosity that
normalises each curve is given in the last column of Table
\ref{haloprop}. The vertical line indicates the virial radius of the halo, while 
the horizontal line shows the surface brightness limit of \citetalias{2008ApJ...681..856R}.
The surface brightness profile shows a central peak
with a tail that flattens out as the column density does the same. 
The emission is more extended in the stronger wind
models. There are two reasons for this. The lower central column density 
means that photons are in general closer to line centre when they 
encounter the outer parts of the halo, where higher wind velocities 
can shift the photons sufficiently back towards line centre to be
scattered by rather low column density gas. Comparing the radial
profile of \hi column density and \lya surface brightness for the
different wind models suggest that this is more important 
than the the \hi column density in the outer parts of the halo.

We have not plotted the (\hi weighted) temperature profile, 
which for all models decreases gently with radius from around
$10^{5}$K to $10^{4.3}$ K. 
The central temperature is slightly higher in the weak wind models.

We can now compare the properties of the haloes to the modelling 
of \citetalias{2010MNRAS.403..870B}. Unlike in our previous 
analytical modelling the column density in our simulations here 
has a pronounced core at very small radii. At larger radii the 
radial profile of the \hi column density in the MDW model appears
rather similar even though much larger samples of simulated haloes 
would be needed to make a more meaningful 
comparison; we refer the reader to
\citet{2009MNRAS.397..411T} for a quantitative comparison with the observed DLA column
density distribution. The radial velocity profiles in our full
cosmological simulations are obviously much more complex than
the simple power law profiles we had assumed in our previous work, 
although there is a slight trend toward lower velocities at the
centre of the halo. There was obviously also no possibility 
to account for the 3D effects of any clumping in our previous 1D
modelling.


\subsection{2D Views of the Halo} \label{S:2Dview}

Figure \ref{fig:gr2xm} shows maps of the \hi column density 
distribution and \lya surface brightness and 2D spectra 
for Halo 2 for the MDW model looking along
the $x$-axis from $-\infty$ (left) and $+\infty$ (right). Note that
some of the axes have been flipped so that the left and right columns
have their axes pointing in the same direction. The dashed black curve
shows the virial radius.

\pmcomp{momentum}{7}{1}{2D images and spectra for Halo 2 in the MDW
simulation. The left (right) column is as viewed from $x = - \infty ~ (+ \infty)$. 
The upper two panels show the \hi column density for the \emph{closer 
half} of the simulation box. The dashed circle shows the virial radius. 
The contour represents $N_\ro{DLA} = 10^{20.3}
\cm$, for a sightline passing through the entire box. The next two panels show the observed \lya 
surface brightness. The contour shows the surface brightness limit of the
\citetalias{2008ApJ...681..856R} survey, $S_0 = 10^{-19}$ erg/s/cm$^2$/arcsec$^2$. 
The lower four panels show 2D spectra, calculated using a ``slit''
placed along either the $y$- or $z$-direction (labelled in the
vertical axis) and with a width of 2 arcsec, which is the same width
as in \citet{2008ApJ...681..856R}. The contour shows the spectral
intensity limit of the \citetalias{2008ApJ...681..856R} survey (Rauch, M., private
communication), which is $S_\lambda \approx 4 \times 10^{-20}$
erg/s/cm$^2$/arcsec$^2$/\AA. The Figure is discussed further in 
Section \ref{S:2Dview}.}{fig:gr2xm}

The top panels show the \hi column density along a line of sight parallel
to the $x$-axis. Each panel corresponds to a line of sight that passes
through the \emph{nearer half of the box}. In other words, the colours in the
legend to the right of each Figure represent the column density of
neutral hydrogen that the photon (emitted at the centre of the halo)
would have to  pass through if it were  to reach the observer directly
without any scattering.
The solid line in
the top two panels is a contour representing $N_\ro{DLA} = 10^{20.3}
\cm$, calculated for sightlines passing through the \emph{entire} box;
it is thus the same in the left and right top panels. A sightline
passing through the region inside the contour would encounter a DLA.

The column density distribution reflects the irregular distribution of gas, with a
mixture of thin filaments and relatively isolated clumps of neutral
hydrogen. The differences between the panels show 
the asymmetry in the distribution of gas. 
The cross-section for damped absorption is composed of a
large central clump accompanied by a number of isolated clumps, 
scattered around the halo. The column density of individual 
clumps of neutral hydrogen
drops rapidly below $\sim 10^{17}\cm$, due to the effect of
the ionising UV background on particles below the `cold cloud' 
threshold.

The next two panels show  an image of the halo that is coloured
according to the surface brightness of the \lya emission, calculated\footnote{See
equation (20) of \citet{2009ApJ...696..853L}.} assuming the luminosity
as described previously and given in Table \ref{haloprop}. The solid
contour represents the surface brightness limit of the 
\citetalias{2008ApJ...681..856R} survey, $S_0 = 10^{-19}$ erg/s/cm$^2$/arcsec$^2$.

The surface brightness maps show that the \lya emission generally
traces but is less clumpy than the underlying neutral hydrogen
distribution. In particular, the $S_0$ contour encloses a much larger area
than the DLA contour both by being larger and also by having
fewer ``holes''. The \lya photons are scattered effectively out to
large radii and lower column densities. The $S_0$ contour in the 
surface brightness is roughly similar to that of the $N_\hi \sim
10^{18}\cm$ contour\footnote{Varying between 
$\sim 10^{17} - 10^{19}\cm$. Note that this is not the same as the mean 
integrated column density along the final escape flight of the photon, 
which we have also calculated explicitly using our code and which 
can be approximated  analytically by $N_\ro{esc} \approx 10^{20}\cm ~ (v_\ro{peak} / 500 \kmsec)^2$,
where $v_\ro{peak}$ is the peak of the emission spectrum.}
in the \hi column density plot.

Our central \lya source is able to illuminate clouds of
neutral gas that are detached from the central clump. 
The \lya photons tend to find low-density
paths of escape. For example, the photons face a column density 
of \hi which is on average $\sim 7$ times larger in the $-z$ direction than 
in the $+z$ direction, and the corresponding observed \lya flux is on average $\sim 3$ times 
smaller. For the $x$ direction, as shown in the Figure, the difference 
in the observed flux between two opposite viewpoints is around $70\%$.

The bottom four panels are 2D spectra, calculated using a ``slit''
placed along either the $y$- or $z$-direction (labelled in the
vertical axis) and with a width of 2 arcsec, the same width
as used in \citet{2008ApJ...681..856R}. The contour shows the spectral
intensity limit of the \citetalias{2008ApJ...681..856R} survey, $S_\lambda \approx 4 \times 10^{-20}$
erg/s/cm$^2$/arcsec$^2$/\AA.

The 2D spectra show a dominant blue peak in the $-x$ direction, and the 
opposite trend when viewed from $+x$; see the discussion of Figure 
\ref{fig:spectview} below. 
As expected, the separation of the peaks is largest where the
column density is largest. The dominance of one peak over the other tends to 
increase as we look away from the centre of the halo. This is most likely because
the probability of a photon scattering in the outer parts of the halo is 
quite sensitive to the velocity of the gas.

\begin{figure*} \centering
\mnpg{
%
%
\begin{psfrags}%
\psfragscanon%
%
\psfrag{s05}[t][t]{\color[rgb]{0,0,0}\setlength{\tabcolsep}{0pt}\begin{tabular}{c}$v_{ph}$ [km/s]\end{tabular}}%
\psfrag{s06}[b][b]{\color[rgb]{0,0,0}\setlength{\tabcolsep}{0pt}\begin{tabular}{c}P$_{v} \times 10^3$ [km/s]$^{-1}$\end{tabular}}%
\psfrag{s10}[][]{\color[rgb]{0,0,0}\setlength{\tabcolsep}{0pt}\begin{tabular}{c} \end{tabular}}%
\psfrag{s11}[][]{\color[rgb]{0,0,0}\setlength{\tabcolsep}{0pt}\begin{tabular}{c} \end{tabular}}%
\psfrag{s12}[l][l]{\color[rgb]{0,0,0} $z = + \infty$}%
\psfrag{s13}[l][l]{\color[rgb]{0,0,0}Viewpoint}%
\psfrag{s14}[l][l]{\color[rgb]{0,0,0} $x = - \infty$}%
\psfrag{s15}[l][l]{\color[rgb]{0,0,0} $x = + \infty$}%
\psfrag{s16}[l][l]{\color[rgb]{0,0,0} $y = - \infty$}%
\psfrag{s17}[l][l]{\color[rgb]{0,0,0} $y = + \infty$}%
\psfrag{s18}[l][l]{\color[rgb]{0,0,0} $z = - \infty$}%
\psfrag{s19}[l][l]{\color[rgb]{0,0,0} $z = + \infty$}%
%
\psfrag{x01}[t][t]{-500}%
\psfrag{x02}[t][t]{0}%
\psfrag{x03}[t][t]{500}%
%
\psfrag{v01}[r][r]{0}%
\psfrag{v02}[r][r]{1}%
\psfrag{v03}[r][r]{2}%
\psfrag{v04}[r][r]{3}%
\psfrag{v05}[r][r]{4}%
%
\includegraphics[width=\textwidth]{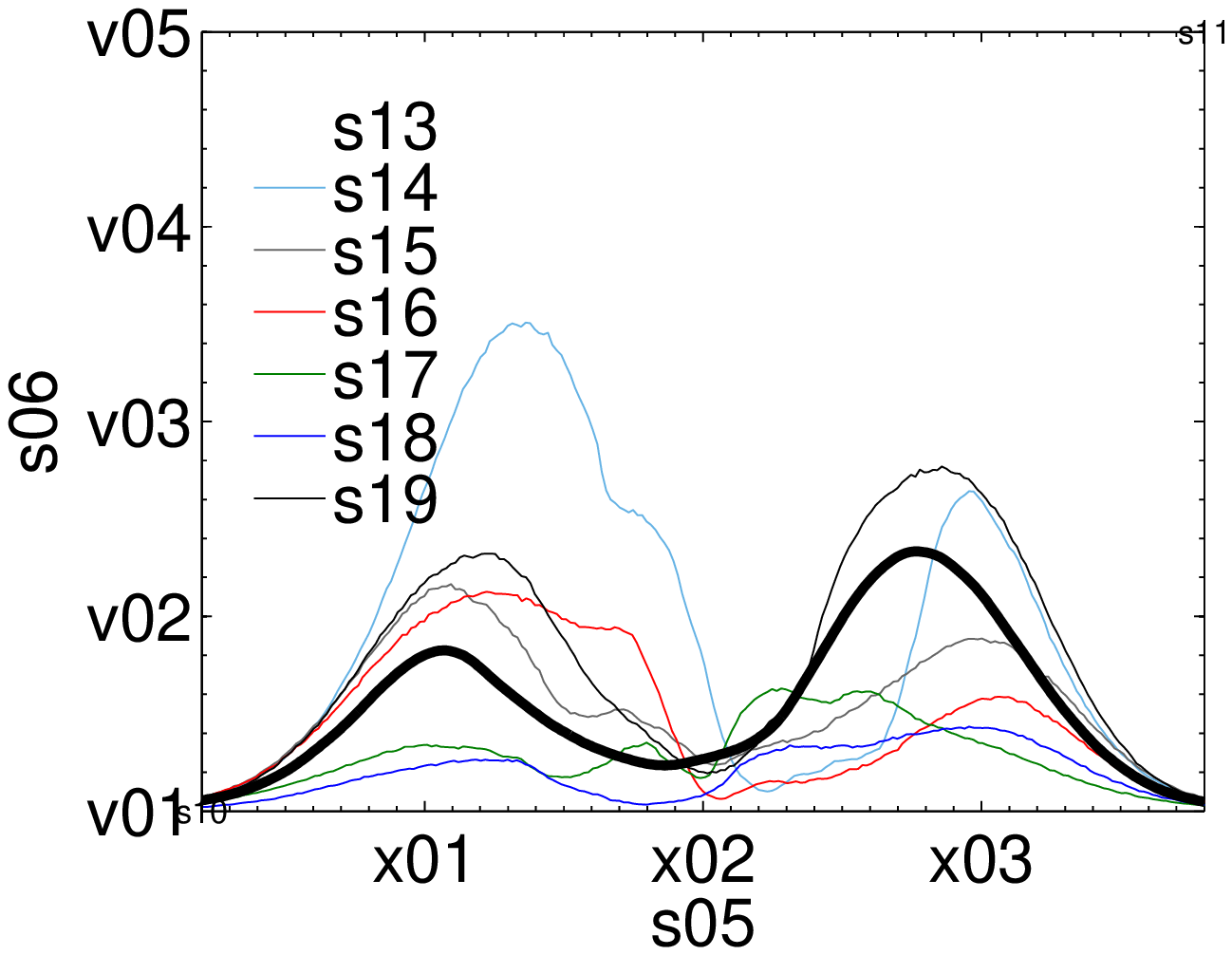}
\end{psfrags}%
%
}
\mnpg{
%
%
\begin{psfrags}%
\psfragscanon%
%
\psfrag{s03}[t][t]{\color[rgb]{0,0,0}\setlength{\tabcolsep}{0pt}\begin{tabular}{c}impact parameter [arcsec]\end{tabular}}%
\psfrag{s04}[b][b]{\color[rgb]{0,0,0}\setlength{\tabcolsep}{0pt}\begin{tabular}{c}log S   [erg/s/cm$^2$/arcsec$^2$]\end{tabular}}%
%
\psfrag{x01}[t][t]{0}%
\psfrag{x02}[t][t]{2}%
\psfrag{x03}[t][t]{4}%
\psfrag{x04}[t][t]{6}%
\psfrag{x05}[t][t]{8}%
%
\psfrag{v01}[r][r]{-22}%
\psfrag{v02}[r][r]{-20}%
\psfrag{v03}[r][r]{-18}%
\psfrag{v04}[r][r]{-16}%
%
\includegraphics[width=\textwidth]{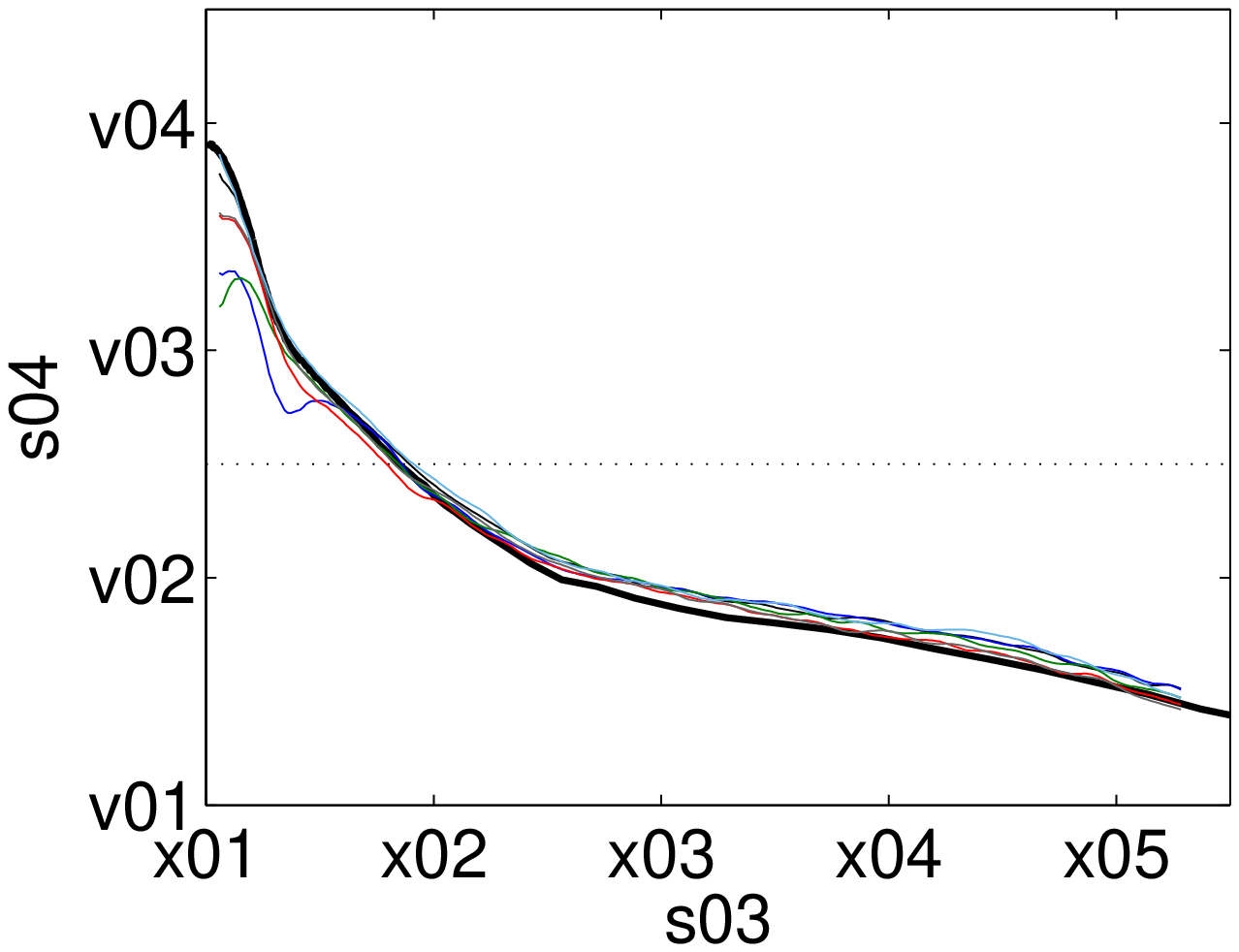}
\end{psfrags}%
%
}
	\caption{The emergent spectrum (left) and surface brightness
profile (right) for Halo 2, showing the effect of the viewing angle. The thick
black line is the angularly-averaged spectrum, which is the same as in
Figure \ref{fig:allplots}. The thin coloured lines show the emergent
spectrum (calculated by collapsing the slit along its spatial
direction, or by averaging over the 2D image) for the 6 different viewpoints as shown in the
legend. The Figure is discussed further in Section \ref{S:2Dview}} \label{fig:spectview}
\end{figure*}

We illustrate the dependence of the emergent spectrum and surface
brightness profile on the viewing angle in Figure \ref{fig:spectview},
which also shows the angularly-averaged value (thick black line). The thin
coloured lines show the emergent spectrum (calculated by collapsing
the slit along its spatial direction, or by averaging over the 2D image) for 6 different
viewpoints along the principal axes. Remember that the observer sees only one of these spectra -
the angularly-averaged quantities are not observable.
The spectra show significant variations, both in which peak dominates and in 
total amount of light observed. The separation of the peaks shows
relatively little variation and the surface brightness profiles are
very similar, especially in the outer parts of the halo.


\subsection{Haloes 1 and 3} \label{S:Halo13}

	\begin{figure*} \centering
		\mnpg{
%
%
\begin{psfrags}%
\psfragscanon%
%
\psfrag{s03}[t][t]{\color[rgb]{0,0,0}\setlength{\tabcolsep}{0pt}\begin{tabular}{c}radius [arcsec]\end{tabular}}%
\psfrag{s04}[b][b]{\color[rgb]{0,0,0}\setlength{\tabcolsep}{0pt}\begin{tabular}{c}radial velocity (vol. wgt.) [km/s]\end{tabular}}%
%
\psfrag{x01}[t][t]{0}%
\psfrag{x02}[t][t]{2}%
\psfrag{x03}[t][t]{4}%
\psfrag{x04}[t][t]{6}%
\psfrag{x05}[t][t]{8}%
%
\psfrag{v01}[r][r]{-100}%
\psfrag{v02}[r][r]{-50}%
\psfrag{v03}[r][r]{0}%
\psfrag{v04}[r][r]{50}%
\psfrag{v05}[r][r]{100}%
\psfrag{v06}[r][r]{150}%
\psfrag{v07}[r][r]{200}%
%
\includegraphics[width=\textwidth]{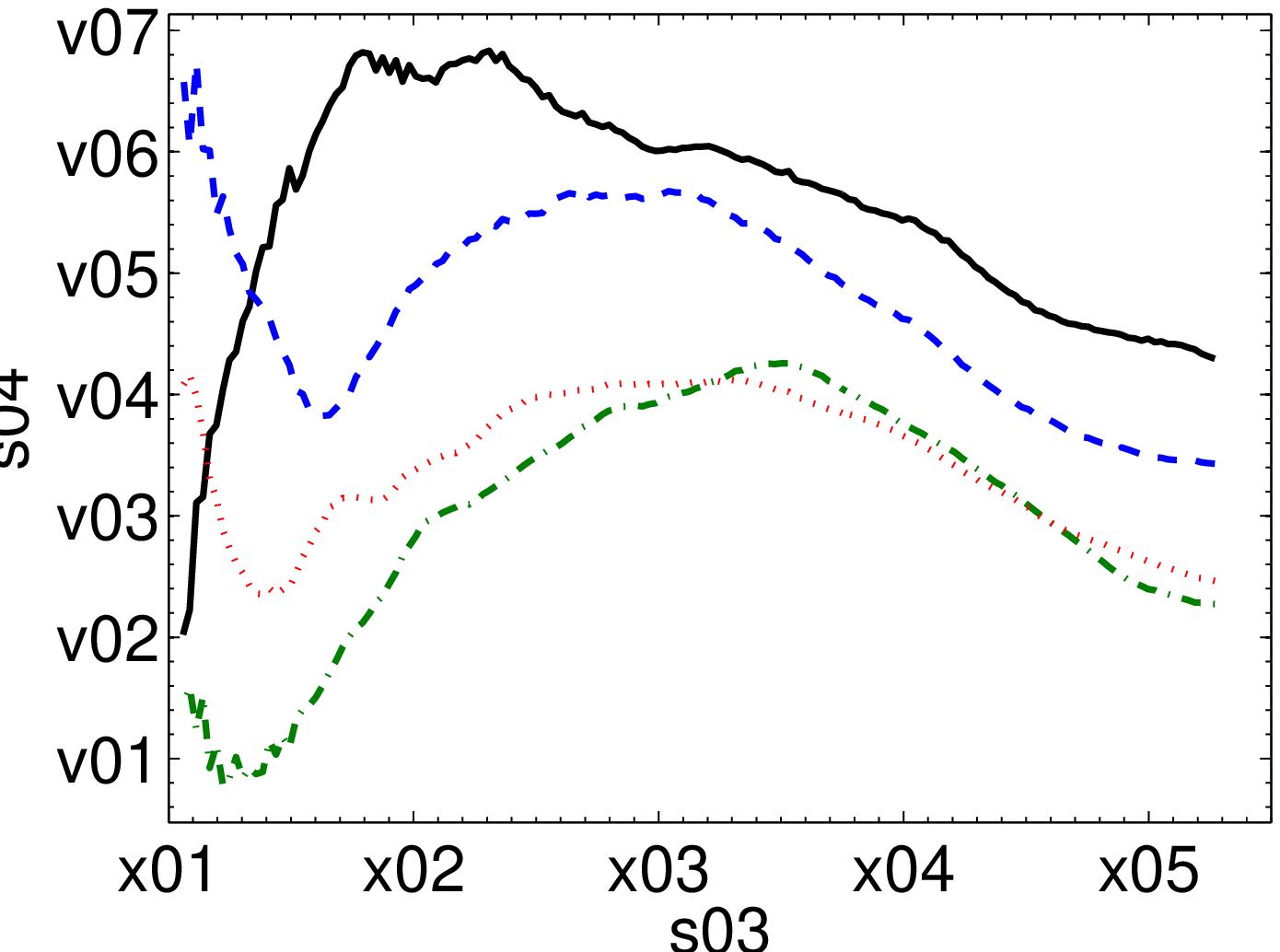}
\end{psfrags}%
%
}\hsp
		\mnpg{
%
%
\begin{psfrags}%
\psfragscanon%
%
\psfrag{s05}[t][t]{\color[rgb]{0,0,0}\setlength{\tabcolsep}{0pt}\begin{tabular}{c}impact parameter [arcsec]\end{tabular}}%
\psfrag{s06}[b][b]{\color[rgb]{0,0,0}\setlength{\tabcolsep}{0pt}\begin{tabular}{c}log HI Column density [cm$^{-2}$]\end{tabular}}%
\psfrag{s10}[][]{\color[rgb]{0,0,0}\setlength{\tabcolsep}{0pt}\begin{tabular}{c} \end{tabular}}%
\psfrag{s11}[][]{\color[rgb]{0,0,0}\setlength{\tabcolsep}{0pt}\begin{tabular}{c} \end{tabular}}%
\psfrag{s12}[l][l]{\color[rgb]{0,0,0}NW}%
\psfrag{s13}[l][l]{\color[rgb]{0,0,0}MDW}%
\psfrag{s14}[l][l]{\color[rgb]{0,0,0}SW}%
\psfrag{s15}[l][l]{\color[rgb]{0,0,0}WW}%
\psfrag{s16}[l][l]{\color[rgb]{0,0,0}NW}%
%
\psfrag{x01}[t][t]{0}%
\psfrag{x02}[t][t]{2}%
\psfrag{x03}[t][t]{4}%
\psfrag{x04}[t][t]{6}%
\psfrag{x05}[t][t]{8}%
%
\psfrag{v01}[r][r]{16}%
\psfrag{v02}[r][r]{17}%
\psfrag{v03}[r][r]{18}%
\psfrag{v04}[r][r]{19}%
\psfrag{v05}[r][r]{20}%
\psfrag{v06}[r][r]{21}%
\psfrag{v07}[r][r]{22}%
%
\includegraphics[width=\textwidth]{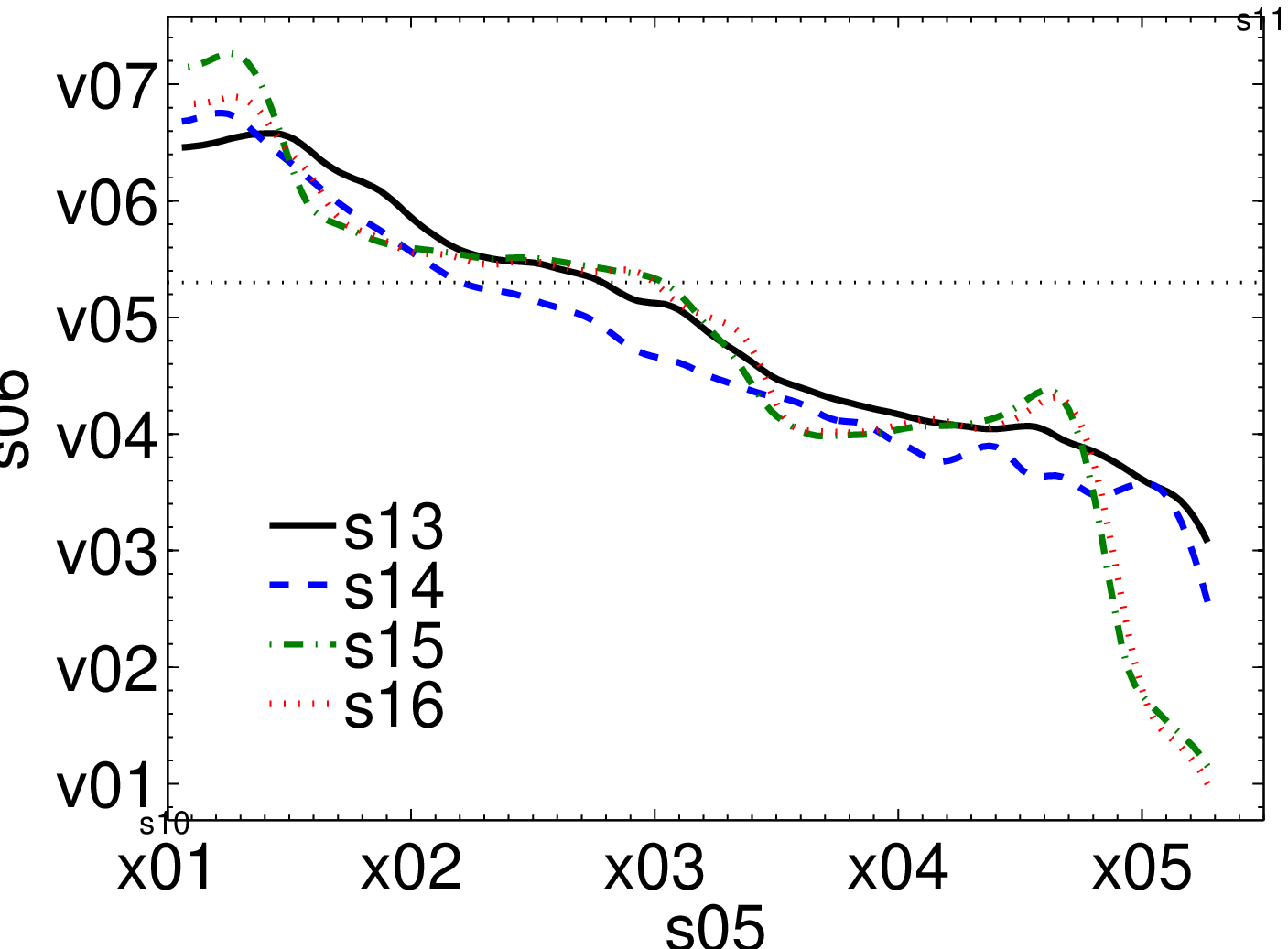}
\end{psfrags}%
%
}
		\mnpg{
%
%
\begin{psfrags}%
\psfragscanon%
%
\psfrag{s03}[t][t]{\color[rgb]{0,0,0}\setlength{\tabcolsep}{0pt}\begin{tabular}{c}$v_{ph}$ [km/s]\end{tabular}}%
\psfrag{s04}[b][b]{\color[rgb]{0,0,0}\setlength{\tabcolsep}{0pt}\begin{tabular}{c}P$_{v} \times 10^3$ [km/s]$^{-1}$\end{tabular}}%
%
\psfrag{x01}[t][t]{-2000}%
\psfrag{x02}[t][t]{-1000}%
\psfrag{x03}[t][t]{0}%
\psfrag{x04}[t][t]{1000}%
\psfrag{x05}[t][t]{2000}%
%
\psfrag{v01}[r][r]{0}%
\psfrag{v02}[r][r]{0.2}%
\psfrag{v03}[r][r]{0.4}%
\psfrag{v04}[r][r]{0.6}%
\psfrag{v05}[r][r]{0.8}%
\psfrag{v06}[r][r]{1}%
%
\includegraphics[width=\textwidth]{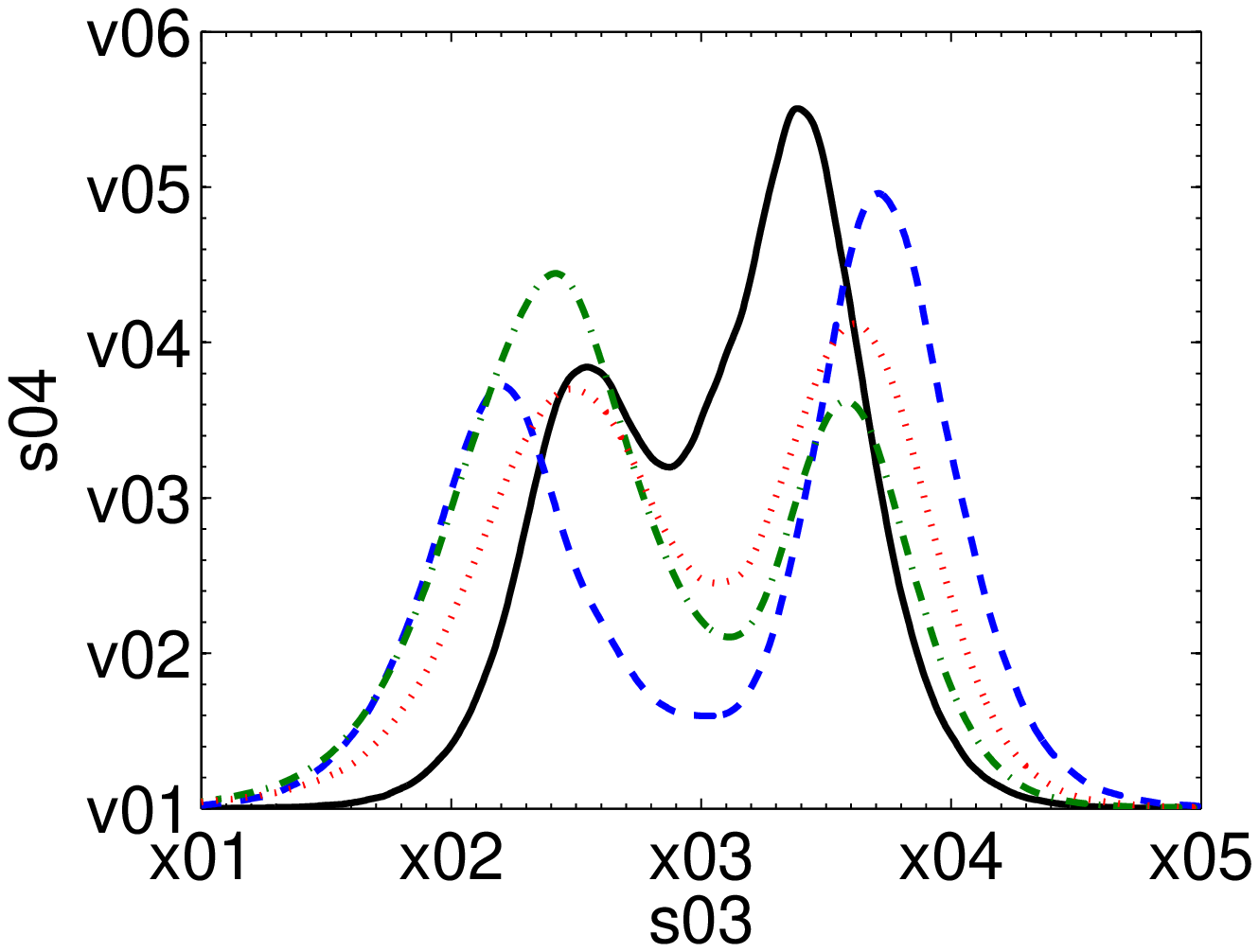}
\end{psfrags}%
%
}\hsp
		\mnpg{
%
%
\begin{psfrags}%
\psfragscanon%
%
\psfrag{s03}[t][t]{\color[rgb]{0,0,0}\setlength{\tabcolsep}{0pt}\begin{tabular}{c}impact parameter [arcsec]\end{tabular}}%
\psfrag{s04}[b][b]{\color[rgb]{0,0,0}\setlength{\tabcolsep}{0pt}\begin{tabular}{c}log S   [erg/s/cm$^2$/arcsec$^2$]\end{tabular}}%
%
\psfrag{x01}[t][t]{0}%
\psfrag{x02}[t][t]{2}%
\psfrag{x03}[t][t]{4}%
\psfrag{x04}[t][t]{6}%
\psfrag{x05}[t][t]{8}%
%
\psfrag{v01}[r][r]{-21}%
\psfrag{v02}[r][r]{-20}%
\psfrag{v03}[r][r]{-19}%
\psfrag{v04}[r][r]{-18}%
\psfrag{v05}[r][r]{-17}%
\psfrag{v06}[r][r]{-16}%
%
\includegraphics[width=\textwidth]{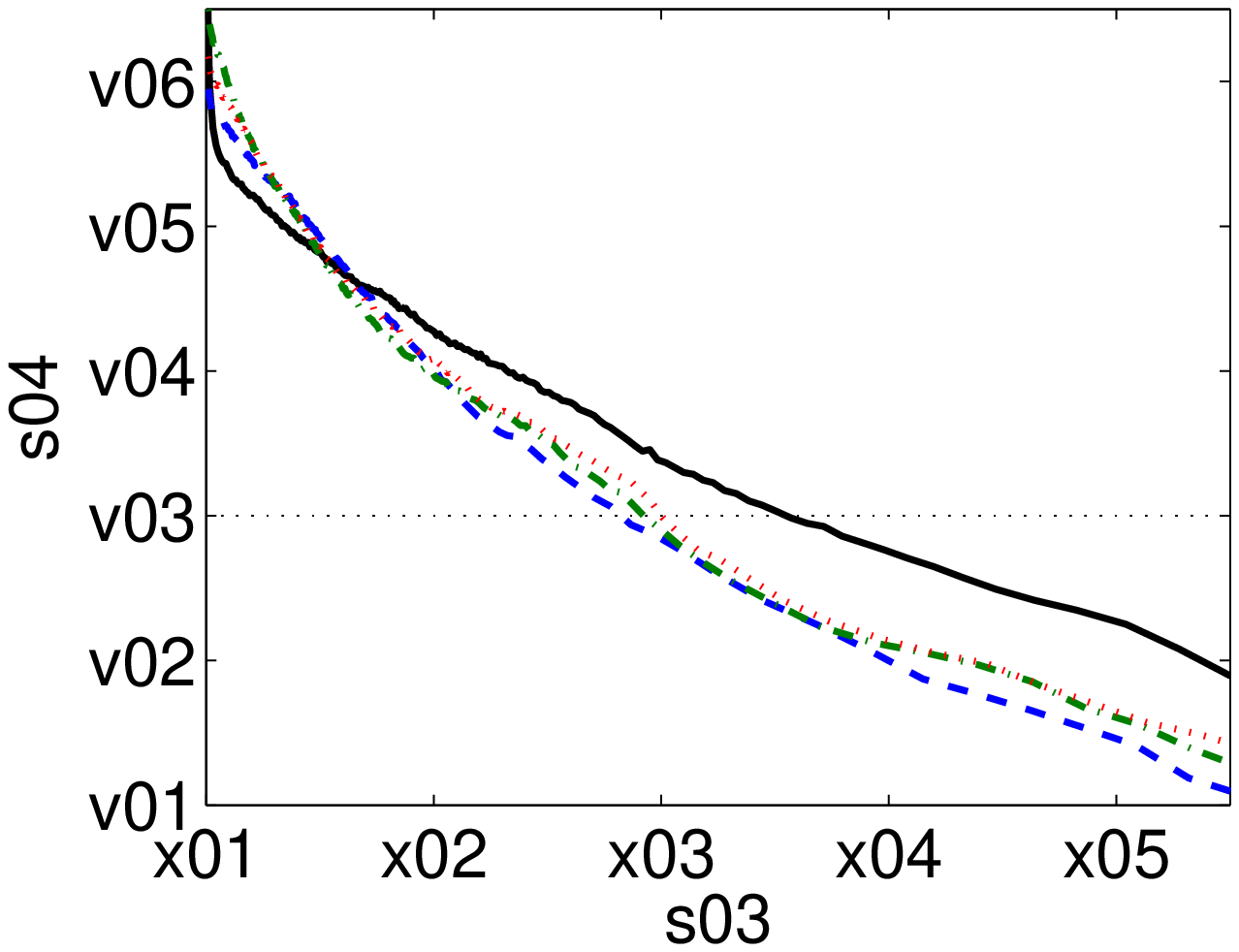}
\end{psfrags}%
%
}
		\caption{Angularly-averaged properties of Halo 1, for the wind
prescriptions shown in the legend (upper left), as discussed in
Section \ref{S:Halo13}. The upper left panel shows the 
volume-averaged radial velocity profile. The upper right panel shows the 
\hi column density as a function of impact parameter. The lower 
left panel shows the spectra, normalised to unit area. The lower right 
panel shows the surface brightness distribution as a function 
of impact parameter.} \label{fig:av1sw}
	\end{figure*}

The angularly-averaged properties of Halo 1 are shown in Figure \ref{fig:av1sw}. 
The panels show the volume-averaged radial velocity profile, 
projected column density of neutral hydrogen, the emergent \lya spectrum and the 
\lya surface brightness as a function of radius/impact parameter. The four 
different wind models are shown on each plot.

The radial velocity profile shows the effect of the outflowing winds in the 
low density gas. The winds provide escape paths filled with gas that is 
outflowing at $\sim 100 - 150$ \kmsec, while in the weaker wind cases the 
average velocity is closer to zero\footnote{The fact that the 
volume-averaged radial velocity is slightly positive in the no-wind simulation 
is most likely due to the centre of mass of the baryons being offset from the centre of mass 
of the halo i.e. the baryons have overshot the bottom of the potential well. Heating 
of the hot, rarefied ISM by the UVB may also contribute.}.

The column density profile of Halo 1 is somewhat smoother than that of
Halo 2 and 3, and as expected the DLA cross section is significantly larger.
Once again we see that the winds have lowered the central column density of the halo.

For Halo 1 the spectra show again the classic double-peaked profile
with a general trend toward  domination of the red peak for haloes with stronger
winds. The surface brightness profiles are rather similar: they peak
at the centre, and decrease smoothly as the column density drops. 
The weaker wind models are as expected more centrally peaked than
the stronger wind models.

	\begin{figure*} \centering
		\mnpg{
%
%
\begin{psfrags}%
\psfragscanon%
%
\psfrag{s03}[t][t]{\color[rgb]{0,0,0}\setlength{\tabcolsep}{0pt}\begin{tabular}{c}radius [arcsec]\end{tabular}}%
\psfrag{s04}[b][b]{\color[rgb]{0,0,0}\setlength{\tabcolsep}{0pt}\begin{tabular}{c}radial velocity (vol. wgt.) [km/s]\end{tabular}}%
%
\psfrag{x01}[t][t]{0}%
\psfrag{x02}[t][t]{0.5}%
\psfrag{x03}[t][t]{1}%
\psfrag{x04}[t][t]{1.5}%
\psfrag{x05}[t][t]{2}%
\psfrag{x06}[t][t]{2.5}%
\psfrag{x07}[t][t]{3}%
\psfrag{x08}[t][t]{3.5}%
%
\psfrag{v01}[r][r]{-100}%
\psfrag{v02}[r][r]{-50}%
\psfrag{v03}[r][r]{0}%
%
\includegraphics[width=\textwidth]{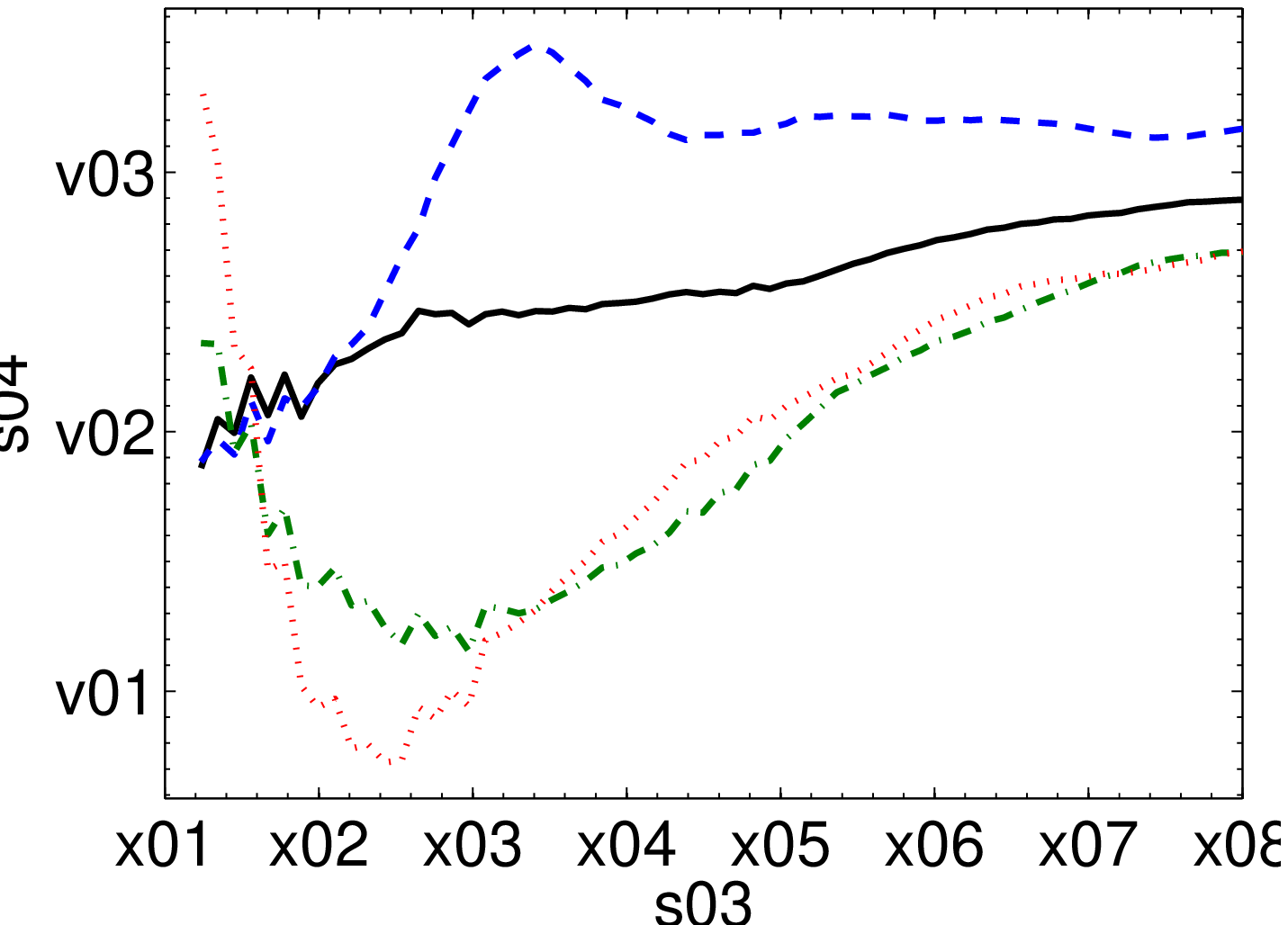}
\end{psfrags}%
%
}\hsp
		\mnpg{
%
%
\begin{psfrags}%
\psfragscanon%
%
\psfrag{s05}[t][t]{\color[rgb]{0,0,0}\setlength{\tabcolsep}{0pt}\begin{tabular}{c}impact parameter [arcsec]\end{tabular}}%
\psfrag{s06}[b][b]{\color[rgb]{0,0,0}\setlength{\tabcolsep}{0pt}\begin{tabular}{c}log HI Column density [cm$^{-2}$]\end{tabular}}%
\psfrag{s10}[][]{\color[rgb]{0,0,0}\setlength{\tabcolsep}{0pt}\begin{tabular}{c} \end{tabular}}%
\psfrag{s11}[][]{\color[rgb]{0,0,0}\setlength{\tabcolsep}{0pt}\begin{tabular}{c} \end{tabular}}%
\psfrag{s12}[l][l]{\color[rgb]{0,0,0}NW}%
\psfrag{s13}[l][l]{\color[rgb]{0,0,0}MDW}%
\psfrag{s14}[l][l]{\color[rgb]{0,0,0}SW}%
\psfrag{s15}[l][l]{\color[rgb]{0,0,0}WW}%
\psfrag{s16}[l][l]{\color[rgb]{0,0,0}NW}%
%
\psfrag{x01}[t][t]{0}%
\psfrag{x02}[t][t]{0.5}%
\psfrag{x03}[t][t]{1}%
\psfrag{x04}[t][t]{1.5}%
\psfrag{x05}[t][t]{2}%
\psfrag{x06}[t][t]{2.5}%
\psfrag{x07}[t][t]{3}%
\psfrag{x08}[t][t]{3.5}%
%
\psfrag{v01}[r][r]{14}%
\psfrag{v02}[r][r]{16}%
\psfrag{v03}[r][r]{18}%
\psfrag{v04}[r][r]{20}%
\psfrag{v05}[r][r]{22}%
%
\includegraphics[width=\textwidth]{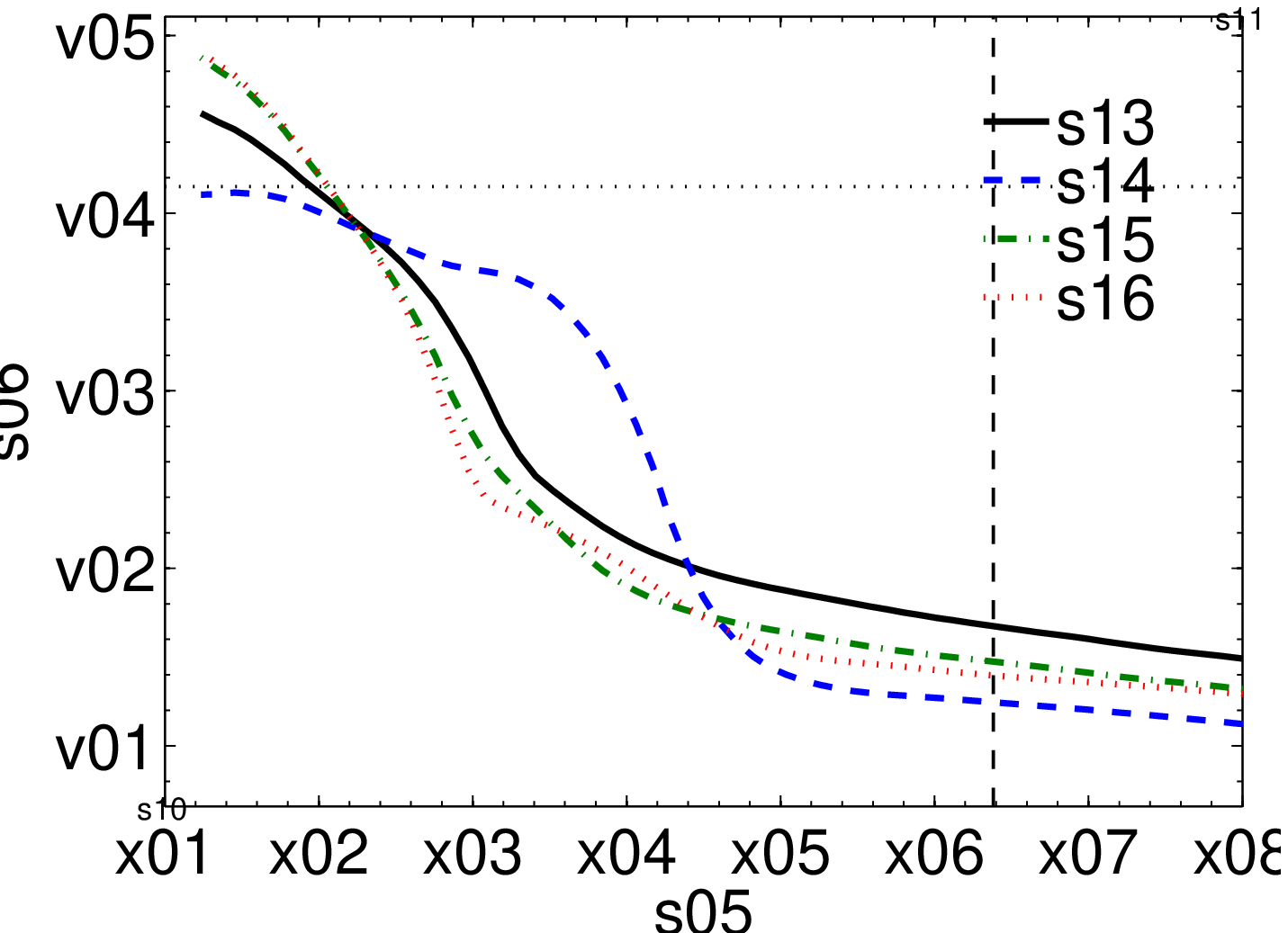}
\end{psfrags}%
%
}
		\mnpg{
%
%
\begin{psfrags}%
\psfragscanon%
%
\psfrag{s03}[t][t]{\color[rgb]{0,0,0}\setlength{\tabcolsep}{0pt}\begin{tabular}{c}$v_{ph}$ [km/s]\end{tabular}}%
\psfrag{s04}[b][b]{\color[rgb]{0,0,0}\setlength{\tabcolsep}{0pt}\begin{tabular}{c}P$_{v} \times 10^3$ [km/s]$^{-1}$\end{tabular}}%
%
\psfrag{x01}[t][t]{-1000}%
\psfrag{x02}[t][t]{-500}%
\psfrag{x03}[t][t]{0}%
\psfrag{x04}[t][t]{500}%
\psfrag{x05}[t][t]{1000}%
%
\psfrag{v01}[r][r]{0}%
\psfrag{v02}[r][r]{2}%
\psfrag{v03}[r][r]{4}%
\psfrag{v04}[r][r]{6}%
%
\includegraphics[width=\textwidth]{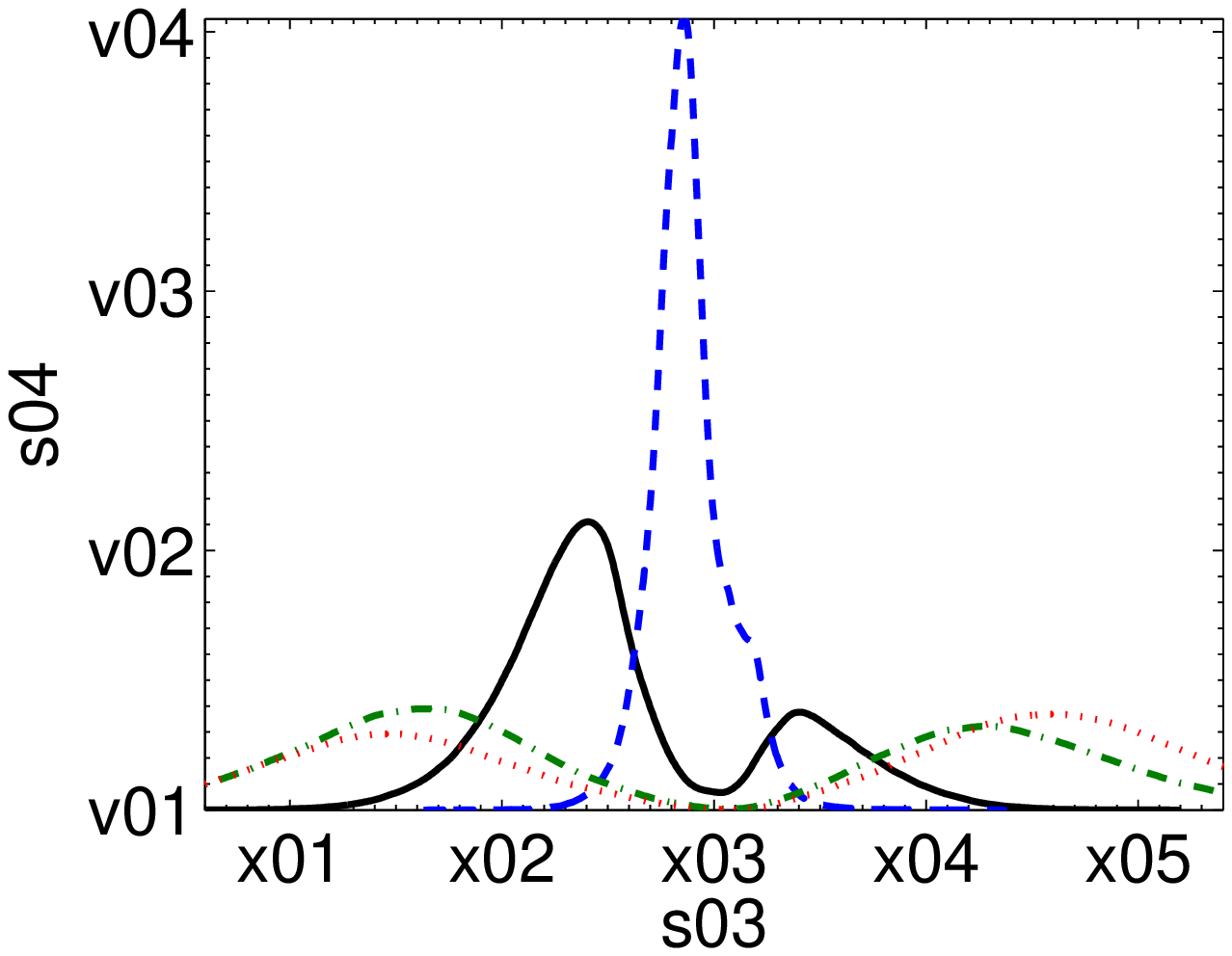}
\end{psfrags}%
%
}\hsp
		\mnpg{
%
%
\begin{psfrags}%
\psfragscanon%
%
\psfrag{s03}[t][t]{\color[rgb]{0,0,0}\setlength{\tabcolsep}{0pt}\begin{tabular}{c}impact parameter [arcsec]\end{tabular}}%
\psfrag{s04}[b][b]{\color[rgb]{0,0,0}\setlength{\tabcolsep}{0pt}\begin{tabular}{c}log S   [erg/s/cm$^2$/arcsec$^2$]\end{tabular}}%
%
\psfrag{x01}[t][t]{0}%
\psfrag{x02}[t][t]{1}%
\psfrag{x03}[t][t]{2}%
\psfrag{x04}[t][t]{3}%
%
\psfrag{v01}[r][r]{-24}%
\psfrag{v02}[r][r]{-22}%
\psfrag{v03}[r][r]{-20}%
\psfrag{v04}[r][r]{-18}%
\psfrag{v05}[r][r]{-16}%
%
\includegraphics[width=\textwidth]{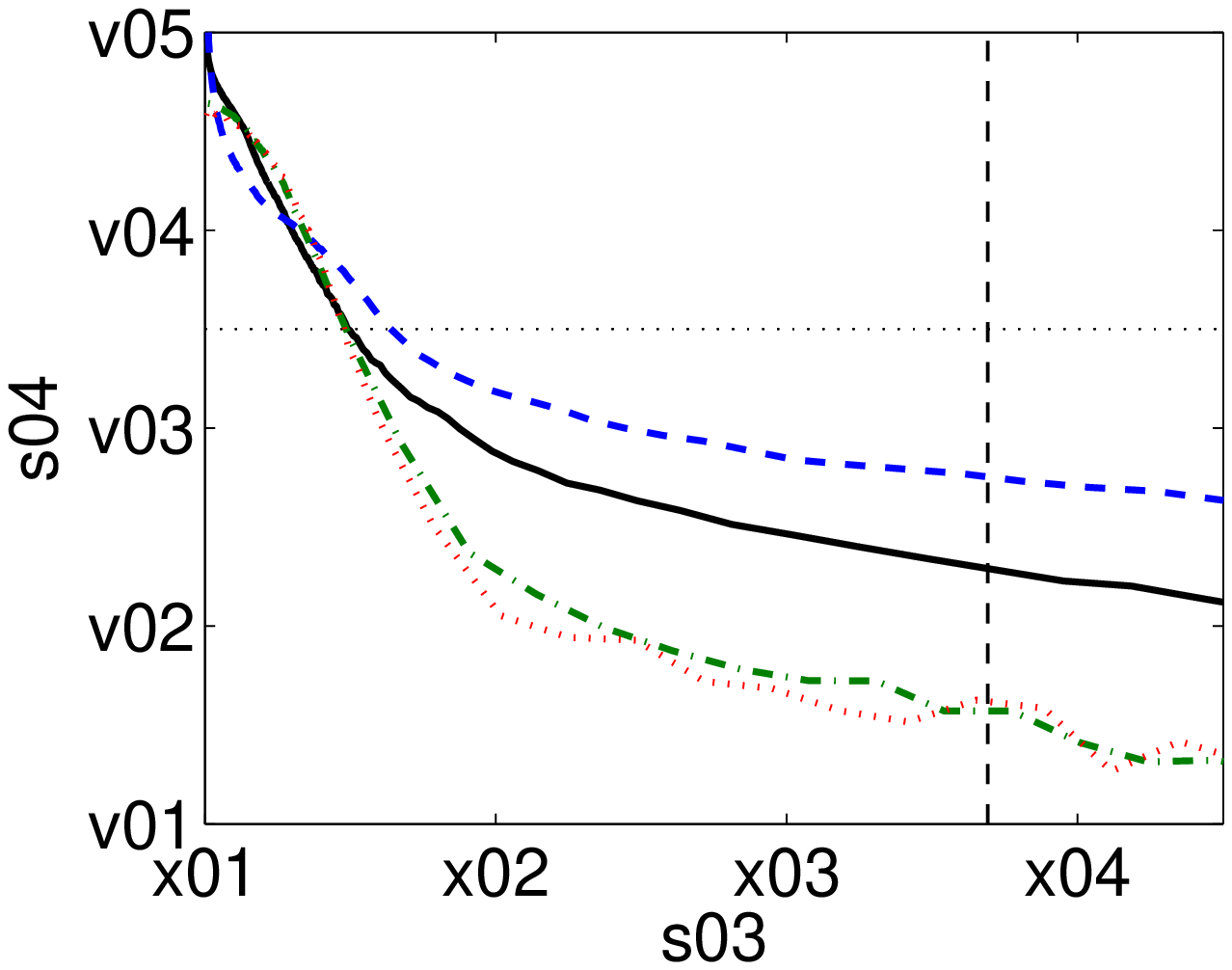}
\end{psfrags}%
%
}
		\caption{Angularly-averaged properties of Halo 3, for the wind
prescriptions shown in the legend (upper left), as discussed in 
Section \ref{S:Halo13}. The panels plot the same quantities as 
in Figure \ref{fig:av1sw}.} \label{fig:av3sw}
	\end{figure*}

The angularly-averaged properties of Halo 3 are shown in Figure \ref{fig:av3sw}. 
The gas in the stronger wind models is more extended. Volume-weighted all of the 
models are dominated by infall at small radii, with only the 
SW model showing a very mild outflow at large radii. The MDW wind clearly 
decreases the average infall rate, but not to the extent that the infall 
is reversed.

The angularly-averaged spectrum of Halo 3 is a particularly nice illustration of
inflowing gas at the centre of the halo producing a more 
prominent blue peak in the spectrum. In
spite of this, there are viewing angles  for which the spectrum is
dominated by a red peak --- see Figure \ref{fig:wc3xm} below. 
The surface density profile reflects the
column density profile --- the inner, dense core results in a central
peak whose rapid decline gives way to a shallower plateau as we reach
the outer, diffuse background.

Figures \ref{fig:wc1xm}, \ref{fig:wc2xm} and \ref{fig:wc3xm} show a
comparison of the 2D maps and spectra for the MDW (left) and WW
(right) wind prescriptions for the three haloes.
Note that the left panel of Figure \ref{fig:wc2xm} is the same as the left
panel of Figure \ref{fig:gr2xm}, and is reproduced in this Figure for
convenience. As noted above, the emission is more extended in the stronger wind
models, and stronger winds produce a narrower spectral distribution (note the 
difference in the horizontal scale of the 2D spectra).

\windcomp{momentum}{weak}{0}{1}{1}{2D images and spectra for the MDW
(left) and WW (right) wind prescriptions for Halo 1 viewed from $x =
-\infty$. The panels are the same as in Figure \ref{fig:gr2xm}. This 
Figure is discussed further in Section \ref{S:Halo13}.}{fig:wc1xm}

\windcomp{momentum}{weak}{7}{1}{1}{2D images and spectra for the MDW
(left) and WW (right) wind prescriptions for Halo 2 viewed from $x =
-\infty$. The panels are the same as in Figure \ref{fig:gr2xm}. This 
Figure is discussed further in Section \ref{S:Halo13}.}{fig:wc2xm}

\windcomp{momentum}{weak}{4}{1}{1}{2D images and spectra for the MDW
(left) and WW (right) wind prescriptions for Halo 3 viewed from $x =
-\infty$. The panels are the same as in Figure \ref{fig:gr2xm}. This 
Figure is discussed further in Section \ref{S:Halo13}.}{fig:wc3xm}


\section{Discussion} \label{S:discuss}

The dependence of the spectra, in both the line shape and the total
observed surface brightness, on the viewing angle is very striking
and is obviously very important for the interpretation of observed 
(spatially extended low surface brightness) \lya emitters. 
It suggests, {\it e.g.}, that the large variety of line 
shapes in the observed sample of \citetalias{2008ApJ...681..856R} 
is at least partially an
orientation effect. The dependence of surface brightness on orientation 
further suggest that the ``duty cycle'' that we introduced in 
\citetalias{2010MNRAS.403..870B} may also be interpreted as an orientation
effect, with only a certain fraction of haloes seen at an
orientation where the observed flux exceeds the detection 
limit. A rather large sample of haloes would be required to
determine the probability distribution of the 
observed \lya flux from a halo of given mass and intrinsic luminosity.
As briefly mentioned but not investigated in detail here, there should,
however, obviously  be also  a variation of the  spectral shape and
surface brightness distribution  with time in particular when a galaxy
undergoes a major merger and/or starburst. 

\subsection{Limitations of the modelling} \label{S:limit}

We will discuss now again in more detail some of the limitations of our modelling. 
Firstly, the failure of the simulations of
\citet{2009MNRAS.397..411T} to reproduce the high velocity tail 
of the velocity width distribution of the associated metal absorption
in DLAs suggests that higher outflow velocities, a 
more extended spatial distribution of the neutral gas, \emph{and/or} 
a less clumpy and more spatially extended distribution of 
low-ionisation metals is required from our galactic wind implementations. 
The effect of higher outflow velocities on the \lya emission depends 
on exactly how the velocity would be increased. If the bulk velocity in the
centre of the halo is increased, then this would accelerate frequency-space diffusion
resulting in a more centrally peaked surface brightness profile. On
the other hand, if the bulk velocity is increased primarily in the
outer parts of the halo, then this would have a minimal effect,
perhaps shifting photons back toward line-centre and thereby
increasing the apparent size of the emitter. 

We have also not attempted to model the effects of dust on the \lya radiation. 
DLAs are known to be among the most metal-poor systems in the high redshift 
universe, and are thus expected to be relatively dust free 
\citep{2009MNRAS.393..557P}, apart from perhaps the most metal-rich
systems \citep{2010MNRAS.408.2128F,2010arXiv1011.5312F}.
The long random walk of a typical \lya photon makes them particularly 
susceptible to even a small amount of dust. 

Another worry could be our use of a rather coarse regular grid. 
Small-scale inhomogeneities in the \hi density are, however, not likely
to have much effect in the absence of dust. As we have seen the
\lya emission profile smoothes them over. At most, an irregular edge
of an \hi region may lead to a shallower surface brightness
profile. Inhomogeneities in the bulk velocity would have a somewhat greater
effect, increasing the frequency diffusion. These inhomogeneities
could be modelled as a turbulent velocity, which has a similarly 
small effect as raising the temperature.

The fact that we have simply injected \lya photons at the centre
of the halo is a rather strong albeit observationally motivated 
assumption.
\citet{2006ApJ...652..981W} placed limits on the spatial 
extent of in situ star formation in DLAs, concluding that extended sources 
(radius $\gtrsim$ 3kpc, 0.4 arcsec) were very rare\footnote{Some extended star-formation 
has recently been detected \citep{2010arXiv1011.6390R} --- as discussed previously, 
we will investigate the effect of spatially extended \lya sources in future
work.}. As \citetalias{2008ApJ...681..856R} and 
\citetalias{2009MNRAS.397..511B} have argued, \lya emission in the systems we 
are considering is most likely to be powered by star 
formation, suggesting that their large observed sizes are the result of 
radiative transfer through the surrounding neutral gas. Our main focus 
here has been on whether such a scenario is indeed plausible. 
The fact that the
\hi density peaks very near the centre of the halo means that the
majority of photons should be emitted near the centre of the halo, as
we have assumed. Making the intrinsic \lya emission more extended
would obviously make the observed emission also more extended.
A more realistic \lya emissivity would need to consider
the interaction of UV radiation from star formation with the
surrounding \hi, as well as the possible contribution from 
cooling radiation \citep{2010ApJ...725..633F} and fluorescence 
due to the UV background and quasars \citep{2005ApJ...628...61C,2008ApJ...672...48C}. 
However, fluorescence is expected to contribute little to the \lya emission 
from DLAs (\citetalias{2008ApJ...681..856R}), and cooling is likely to only 
be significant in massive haloes, which
aren't expected to host the bulk of the DLA population. 
While cooling radiation may not account for the bulk of DLA lya
luminosity, it could still be a relevant factor in 
determining the observed size of the emitters, as the surface
brightness of the cooling radiation may drop less rapidly towards large
radii than the radiation scattered outwards from central star-forming
regions \citep{2006ApJ...649...14D,2009MNRAS.400.1109D,2010ApJ...725..633F}.

We have considered here a rather small sample of haloes, for several
reasons. Firstly, it allows us to discuss some of the 
many factors relevant for the complicated interpretation 
of the \lya emission at length.
Also, the simulations of \citet{2009MNRAS.397..411T} used a box
size that was too small to
produce large numbers of haloes with $M_v > 10^{10.5} \Msol$. This was
necessary to give an acceptable resolution. Our work here
should thus be seen as complementary to the work of \citetalias{2010MNRAS.403..870B}. 
Many of the assumptions made there are borne out in the simulations 
described here. The \lya emitters have considerable spatial extent, and
are larger than the corresponding DLA cross-section as \lya photons
can be effectively scattered by gas with column densities  less
than that of a DLA; the spectra are often double-peaked, with
inflow/outflow producing a dominant blue/red peak; the radial velocity
and mass flow rate are quite moderate toward the centre of the halo,
increasing the spatial extent of the \lya emission. 

\subsection{Relation to LBGs}

Observationally well studied LBGs at similar redshift -- presumably the more luminous
and more massive cousins of typical DLA host galaxies -- offer the
opportunity to investigate their \lya emission in considerably more
detail, though a comparison may be hampered due to the increased
role of dust and the fact that galactic winds are likely to be more
powerful in these rapidly star-forming galaxies. \citet{2010ApJ...717..289S} have
summarized the properties of the observed \lya emission in LBGs.
\lya is seen in emission and absorption with a wide range of EW. 
When observed in emission, \lya typically shows a dominant 
red peak with offsets in the range of 200-800 \kmsec with respect to
the systemic redshift as defined by the nebular emission lines.
The medium offset of the \lya emission is $485 \pm 175$ \kmsec
with a typical width of 500 \kmsec \citep[e.g.][]{2009MNRAS.398.1263Q}. 
While the \lya emission is offset
towards the red, the interstellar absorption lines are offset to the
blue with a mean of about $-250$ \kmsec and a tail of absorption 
often extending out to $-800$ \kmsec. In particular, the blueshift 
of the interstellar absorption line has been convincingly interpreted
as being due to outflowing gas moving towards the observer \citep{2006A&A...460..397V}. 
The most massive of our three haloes has a mass which is approaching
that inferred for LBGs DM haloes so it is interesting to see
how our simulations compare. It is the momentum wind driven
model with its pronounced red peak which appears to have the most 
similar spectral distribution. The peak of the angular averaged
emission in the MDW model is at about +400 \kmsec. 
The secondary blue peak is, however, considerably stronger than in the composite
spectrum presented in \citet{2010ApJ...717..289S}, perhaps indicating that
the central column density is somewhat too large and that real galactic
winds in LBGs are more efficient in reducing the central optical
depth. In fact, most LBGs do not show an intrinsic damped absorption
system in their spectra and the outflow velocities indicated by the
interstellar absorption lines appear to be higher than that of
the outflowing gas in our simulations.

\subsection{A more detailed comparison with the Rauch et al. emitters}

We come now to a more detailed comparison to 
the \citetalias{2008ApJ...681..856R} emitters. 
As discussed in \citetalias{2008ApJ...681..856R} (see also 
\citetalias{2010MNRAS.403..870B}), for 12 of the 27 spectra only a single emission 
peak is visible while six/three of the spectra show a weak secondary blue/red counter-peak. 
The remaining spectra are extended in frequency space without a clear peak
structure. The widths of the spectral peaks ranges from $\sim 250 -
1000 \kmsec$. The surface brightness profiles are predominantly
centrally peaked with wings that often extend well beyond the Gaussian
core of the PSF. This is particularly true of the brightest sources, 
while the fainter sources are more difficult to characterise due to the noise.
As discussed in detail in section \ref{S:galwind}, for the weak and no wind model
the spectral distribution shows rather symmetric double-peaked
profiles. The \citetalias{2008ApJ...681..856R} emitters appear instead to be similar to the 
two models with the stronger winds. Of the two models with strong
winds, the MDW model shows the more pronounced preference 
for a dominant red peak and generally the contrast between 
the two peaks is largest in this model. This model is thus
probably the one which comes closest to the observed emitters. 
As discussed, the main difference between the models with the different
wind implementations is due to the reduction of the central column density of neutral 
hydrogen due to the effect of the galactic wind. Note that most of
the \citetalias{2008ApJ...681..856R} emitters are only just above the detection threshold and
offer thus a somewhat  limited dynamic range. A weaker second peak would probably not
be detected in many of the observed single-peak emitters. There is,
however, a noticeable difference: some of the \citetalias{2008ApJ...681..856R}
emitters have more than one peak in the spatial direction suggesting 
a more spatially extended source of ionizing photons perhaps with 
several peaks. As discussed previously, in our simulation 
the photons were all injected at the centre of the DM halo. 
As is apparent from the \hi column density maps, 
the \hi distribution is actually also rather clumpy. In a more
quantitative study of a larger sample of haloes it would thus
certainly be worthwhile to test if the correspondence 
between observation and simulations can be improved by injecting \lya
photons in several \hi density peaks. 


\subsection{Implication for the detectability of faint high-redshift
 \lya emitters}

The velocity shift of \lya emission due to resonant scattering is very important for the
detectability of \lya emitters at high redshift \citep[e.g.][]{2004MNRAS.349.1137S}. 
As hydrogen becomes increasingly neutral at high redshift, the 
optical depth to \lya scattering due to the neutral IGM 
between the emitters and the observers steadily increases. 
As discussed e.g. by \citet{2009arXiv0910.2712Z},  
\citet{2010MNRAS.tmp.1158D}and \citet{2010arXiv1009.1384L}, 
the effective transmission thereby depends sensitively on 
the optical depth to \lya scattering and the details of the gas 
velocities within the emitter and the intervening IGM. 
Velocity shifts of about 500 \kmsec as observed 
in $z \sim 3$ LBG/LAEs would strongly increase the transmission
which for appreciable shifts to the red scales as 
\citep{1998ApJ...501...15M,2007MNRAS.377.1175D,2010MNRAS.tmp.1158D},
$\tau (\Delta v) \approx 2.7 x_{HI} \left( \frac{\Delta v}{300
 \kmsec}\right)^{-1} \left( \frac{1+z}{7} \right)^{3/2}$. Our modelling 
for the much fainter \citetalias{2008ApJ...681..856R} emitters still suggests 
shifts of about 300 \kmsec which is good news for the 
detectability of faint \lya emitters at high redshift 
perhaps even before reionization is complete. Note, however that in
our model the shift towards the red decreases rather than increases
with increasing strength of galactic winds. As discussed in Section 
\ref{S:galwind}, this reduced shift towards the red is due to the decreased central 
column density of neutral hydrogen in the models with the stronger galactic
winds. For a good estimate of the expected velocity shift it should
thus be more important to predict the correct column density than 
the exact outflow velocity of the gas. 


\section{Conclusions} \label{S:conclusion}

We have used here realistic numerical simulations of DLA host galaxies 
including the effect of galactic winds to investigate the spatial
and spectral distribution of low-surface brightness 
\lya emission assuming that the emission is powered by 
star formation at the centre of the corresponding DM haloes.
Our main results can be summarised as follows. 

\begin{itemize}

\item The haloes contain a mixture of inflowing and outflowing gas. As
a result, the angularly-averaged spectrum typically shows two peaks,
with the relative strength of the red (blue) peak being a reflection
of the relative contribution of outflow (inflow) as characterized by 
the volume-weighted radial velocity of \hi. The separation of the two
peaks is mainly governed by the central \hi column density. 
Line-centre photons can escape due to inhomogeneities in the
gas density providing low-density paths of escape and bulk velocities
allowing rapid diffusion in the fluid frame frequency space. 
The different wind implementations lead to significantly
different central \hi column densities as well as different 
radial motions of the gas, which are reflected in 
different spectral shapes of the \lya emission.

\item A comparison of the 2D column density and surface brightness
images shows that the \lya emission region is larger and smoother than
the cross-section for damped absorption. \lya photons escaping at
large radii illuminate regions of protogalaxies that would be probed by absorption 
line spectra with column densities down to $N_\hi \sim 10^{18} \cm$ and below. 
The central source can effectively light up outlying clumps of neutral
hydrogen. Asymmetries in the \hi distribution can dramatically affect
the observed surface brightness, with dense clumps casting shadows as
photons are reflected off their surface. A central source of photons
can illuminate diffuse, highly ionised gas inside and beyond the
virial radius at very faint surface brightness levels.

\item The 2D spectra show considerable variety for the same halo viewed
from different angles. A typical line
profile is double-peaked with one of the peaks dominating. The
separation of the peaks decreases with increasing distance from the
central, dense regions of the halo. The dominance of the red or blue
peak changes with viewing angle, and is affected by the velocity of
the gas relative to the observer. The maximum spectral intensity is
approximately constant across the different haloes at $S_\lambda \approx
10^{-17.5}$ erg/s/cm$^2$/arcsec$^2$/\AA\xspace, with the decrease in
luminosity with mass being countered by reduced spatial and frequency
diffusion.

\item The angularly-averaged surface brightness profile is most
sensitive to the column density of the gas at the centre of the haloes
which in turn is very sensitive to the feedback from galactic
winds. The more efficient feedback implementations result in reduced
column densities at the centre and therefore reduced diffusion in
frequency space and narrower spectral profiles. The maximum blue/redshift 
in our emitters is also primarily sensitive to the central HI 
column density and not the velocity of the wind. For objects with
column densities and wind velocities similar to those in our
simulations, galactic winds may in fact reduce rather than increase 
the visibility of \lya emitters during the epoch of reionization when 
the hydrogen in the IGM is not yet (fully) ionized and \lya emission
from galaxies is strongly suppressed due to the damping wing of the
IGM's Gunn-Peterson trough.

\item For the momentum-driven wind model (and to a somewhat lesser extent 
also the strong wind implementation), our simulated emitters show encouraging 
agreement with the properties of the \citetalias{2008ApJ...681..856R} 
emitters and thus further corroborate the
suggestion that these are the long-sought host population of DLAs. 
This preference for the momentum-driven wind models also appears to result in better
agreement with a range of other observed properties of
high-redshift-galaxies and the IGM \citep{2008MNRAS.387..577O}.
There are, however, at the same time differences between observed and
simulated emitters. The simulated spectra do not show quite the preference 
for a dominant red peak the observed emitters appear to suggest. This is 
partially due to the fact that we have not modeled 
the impact of the intervening Intergalactic Medium which will reduce the observed 
emission in the blue peak relative to the red peak. 
Our investigation nevertheless appears to suggest that somewhat larger
outward motion are required than present in our MDW simulations where 
especially in the smaller haloes the feedback due to the galactic
winds is not quite sufficient to globally reverse the infall of gas. 
This may well be related to the fact that the simulations
also struggle to reproduce the tail of high velocity widths for the
associated low-ionization absorption of DLAs. The existence of several
clumps in the observed \lya emission further suggests that the emissivity of ionizing
photons is more spatially extended than the central injection we
have assumed in our simulations. 
\end{itemize}
 
The study of the effect of galactic winds in a cosmological 
galaxy formation simulation presented here contributes further 
to unravelling the complexities of the \lya emission from high
redshift galaxies. The rather strong dependence
of the \lya emission on the details of the wind implementation should
allow spatially extended low-surface brightness \lya emission 
to be turned into an important diagnostic tool for studying the
internal dynamics of galactic winds as well as their effect on 
the interstellar and circumgalactic medium of high-redshift galaxies.


\section*{Acknowledgments} 

We thank Michael Rauch, George Becker, Sebastiano Cantalupo, Mark Dijkstra, Max
Pettini, Dan Stark, Bob Carswell and Richard Bower for helpful discussions. 
We also thank Luca Tornatore and his modified version of \gad that 
was used to run simulations.

MH was partially supported by STFC grant LGAG 092/ RG43335. LAB was partially supported 
by an Overseas Research Scholarship and the Cambridge Commonwealth Trust. MV is supported 
by PRIN/INAF, ASI/AAE grant, INFN PD/51 and by the ERC Starting Grant ``cosmoIGM''.
ET acknowledges a fellowship from the European Commission's
Framework Programme 7, through the Marie Curie Initial
Training Network CosmoComp PITN-GA-2009-238356.

Numerical computations were done on the COSMOS (SGI Altix
3700) supercomputer at DAMTP, the High Performance Computer
Cluster (HPCF) in Cambridge (UK) and at CINECA (Italy).
COSMOS is a UK-CCC facility which is supported by HEFCE,
PPARC and Silicon Graphics/Cray Research. The CINECA (Centro
Interuniversitario del Nord Est per il Calcolo Elettronico) CPU
time has been assigned thanks to an INAF-CINECA grant.

\bsp 
\label{lastpage}
\end{document}